\documentclass[useAMS,usenatbib]{mn2e}
\usepackage{graphicx}
\usepackage{latexsym}
\usepackage{subfigure}
\usepackage{psfig}

\title{Ionisation-induced star formation III: Effects of external triggering on the IMF in clusters}

\author[J. E. Dale, I.A. Bonnell]{J. E. Dale$^{1}$\thanks{E-mail: dale@usm.lmu.de (JED)}, I. A. Bonnell$^{2}$\\
$^{1}$Excellence Cluster `Universe', Boltzmannstr. 2, 85748 Garching, Germany.\\
$^{2}$Department of Physics and Astronomy, University of St Andrews, North Haugh, St Andrews, Fife KY16 9SS}

\begin{document}

\pagerange{\pageref{firstpage}--\pageref{lastpage}} \pubyear{2006}

\maketitle

\label{firstpage}

\def\mnras{MNRAS}
\def\apj{ApJ}
\def\aj{AJ}
\def\aap{A\&A}
\def\apjl{ApJL}
\def\apjs{ApJS}
\def\araa{ARA\&A}
 
\begin{abstract}
We report on Smoothed Particle Hydrodynamics (SPH) simulations of the impact on a turbulent $\sim2\times10^{3}$ M$_{\odot}$ star--forming molecular cloud of irradiation by an external source of ionizing photons. We find that the ionizing radiation has a significant effect on the gas morphology, but a less important role in triggering stars. The rate and morphology of star formation are largely governed by the structure in the gas generated by the turbulent velocity field, and feedback has no discernible effect on the stellar initial mass function. Although many young stars are to be found in dense gas located near an ionization front, most of these objects also form when feedback is absent. Ionization has a stronger effect in diffuse regions of the cloud by sweeping up low--density gas that would not otherwise form stars into gravitationally--unstable clumps. However, even in these regions, dynamical interactions between the stars rapidly erase the correlations between their positions and velocities and that of the ionization front.
\end{abstract}

 
\section{Introduction}
The idea that star formation may be self--propagating or self--triggering is an old one, going back at least to \cite{1977ApJ...214..725E}. There are several other mechanisms acting on a range of scales which may trigger star formation, e.g. the mergers of galaxies \citep[e.g][]{1992ApJ...400..153M}, the passage of gas through galactic spiral arms \citep{1969ApJ...158..123R} or collisions of molecular clouds \citep[e.g][]{1979ApJ...229..578S}, but in this series of papers we focus on the idea that feedback from O--type stars may trigger the formation of stars. Photoionization, winds and supernovae have all been invoked as triggering mechanisms.\\
\indent The collect--and--collapse model, in which a feedback--source generates a gravitationally--unstable shell by sweeping up a uniform cloud has been extensively studied. The model is attractive for its simplicity and has been treated analytically and numerically by several authors \citep{1994A&A...290..421W,2001A&A...374..746W,2007MNRAS.375.1291D}. There is a rapidly--growing body of observational evidence of the collect--and--collapse process at work, e.g. \cite{2006A&A...446..171Z,2006A&A...458..191D,2008A&A...482..585D,2010A&A...518L.101Z} and a consonance between observations and theory is emerging.\\
\indent Star formation in a cloud can also be triggered by the action of stars outside the cloud. In the radiation--driven implosion model, the skin of a molecular cloud is heated by the photoionizing radiation from a nearly OB star and evaporates, driving a reverse shock by the rocket effect \citep{1982ApJ...260..183S,1989ApJ...346..735B}. An otherwise stable clump of gas thus collapses and forms stars. This process has been observed by \cite{1995A&A...301..522L} and by \cite{2003MNRAS.338..545K},  \cite{2009MNRAS.393...21G} and \cite{2010arXiv1007.2727B}.\\
\indent Most systems where triggered star formation has been reported have the complex morphologies expected from the interaction of radiation with a turbulent or inhomogeneous ISM. This problem was first addressed by \cite{1995ApJ...451..675E} and has been studied more recently by \cite{2007MNRAS.377..535D}, \cite{2009ApJ...694L..26G} and \cite{2010arXiv1009.0011G}. Systems of this nature are extremely complicated and thus difficult to interpret observationally. Most authors rely on the geometrical coincidence of  young stellar objects (YSOs) with features such as ionization fronts, bright--rimmed clouds or swept--up gas to detect triggering. \cite{2003ApJ...595..900K} and \cite{2008ApJ...688.1142K} studied the W5 HII region at a variety of wavelengths and assessed the degree of triggering by comparing the locations of YSO's to the ionization sources and the walls of the cavities evacuated by HII regions. In their study of NGC 2467, \cite{2009ApJ...700..506S} show that a much higher fraction of YSOs lie within a projected distance of 0.5 pc from an ionization front than would results from random alignment, and conclude that a substantial fraction of the star formation in this region has been triggered. \cite{2009A&A...503..107P} infer triggering from the overlap of many YSOs with dense molecular gas at the borders of the HII region Sh2-254, although they also find the puzzling result that many young objects are projected to lie within Dolidze 25, the ionizing cluster of this region. \cite{2009A&A...497..789U} observe 45 bright--rimmed clouds and classify them as triggered or untriggered on the basis of whether or not they are being photoionized. While all the criteria used by these authors are certainly suggestive of triggered star formation, the evidence presented in all cases is somewhat circumstantial. The papers in this series are an effort to improve the theoretical understanding of triggered star formation and to place such observational interpretations on firmer ground.\\
\indent In a previous paper \citep{2007MNRAS.377..535D} we distinguished weak triggering (`accelerated star formation'), in which stars that would have formed in the absence of feedback are caused to form earlier, and strong triggering in which feedback causes the birth of stars that would not otherwise exist. We consider strong triggering to be more interesting, since it increases the star formation efficiency. Observationally, it is very difficult to distinguish between these possibilities -- even in the case of the collect--and--collapse process, it is not easy to tell whether a given density enhancement has formed by fragmentation of the shell, or is a pre--existing object being overrun. We demonstrated that external irradiation could produce strong triggering by comparison with the evolution of a cloud evolving in the absence of feedback. However, we found that it was difficult to distinguish the triggered and spontaneous objects simply by observing the cloud. Triggering increased the star formation efficiency in our model cloud both by causing the formation of extra stars and by increasing the masses of spontaneously--forming objects. We therefore speculated (as have several other authors, e.g \cite{1994A&A...290..421W}) that triggering may have some observable effect on the stellar IMF which could be used by observers. However, we were only able to form a small number of objects and our resolution was insufficient to follow the formation of individual stars.\\
\indent In this paper, we perform similar calculations, but we simulate a lower--mass cloud at higher resolution, so that we can construct stellar IMFs from our results and see if they are affected by triggering. Once again, we determine the impact of triggering counterfactually by allowing an identical copy of the same cloud to evolve in the absence of ionizing sources so that we may directly compare the evolution of the same gas and stars. In Sections 2 and 3 we discuss our numerical methods and initial conditions. Section 4 contains our results and Sections 5 and 6 contain our discussion and conclusions respectively.\\
\section{Numerical Methods}
We make use of a hybrid Smoothed Particle Hydrodynamics (SPH)/N--body code in which gas is represented by discrete particles and hydrodynamical forces are computed using the SPH formalism \citep{1992ARA&A..30..543M}, stars are represented by sink partlcles \citep{1995MNRAS.277..362B} and gravitational forces are computed using a binary tree (in the case of the gas particles) or by direct summation (for the sink particles). Formation of sink particles and the subsequent accretion of gas onto them is modelled in the manner described in \cite{1995MNRAS.277..362B}. We use the standard artificial viscosity prescription with $(\alpha, \beta)=(1,2)$\\
\indent We use a modified Larson equation of state, given by
\begin{equation}
P = k \rho^{\gamma}
\end{equation}
where
\begin{equation}
\begin{array}{rlrl}
\gamma  &=  0.75  ; & \hfill &\rho \le \rho_1 \\
\gamma  &=  1.0  ; & \rho_1 \le & \rho  \le \rho_2 \\
\gamma  &=  1.4  ; & \hfill \rho_2 \le &\rho \le \rho_3 \\
\gamma  &=  1.0  ; & \hfill &\rho \ge \rho_3, \\
\end{array}
\end{equation}
and $\rho_1= 5.5 \times 10^{-19} {\rm g\ cm}^{-3} , \rho_2=5.5 \times10^{-15} {\rm g\ cm}^{-3} , \rho_3=2 \times 10^{-13} {\rm g\ cm}^{-3}$. The effective cooling at low density mimics line cooling and ensures that the Jeans mass at the point of fragmentation returns approximately a characteristic stellar mass of $0.5$M$_{\odot}$ \citep{2005A&A...435..611J,2006MNRAS.368.1296B}. The $\gamma=1.0$ segment approximates the dust cooling whilst the $\gamma=1.4$ segment mimics the regime in which the collapsing core becomes optically thick and behaves adiabatically. The final isothermal segment allows extremely high density gas to form sink particles, if it has not done so already.\\ 
\indent We use the method described in \cite{2007MNRAS.382.1759D} to model photoionizing radiation from point sources. We take the source to emit ionizing photons isotropically and use a Str\"omgren integral technique to compute the flux of ionizing photons received by a given SPH particle in a given time interval, allowing us to take into account the time required to ionize the particle. We make use of the on--the--spot approximation, neglecting the diffuse ionizing field. In these simulations, the ionizing source is not a sink particle but an artificial source with a fixed luminosity at a constant position outside the molecular cloud under study.\\
\indent The problem we wish to simulate is that of a turbulent molecular cloud illuminated by an external ionizing source. The problem of the irradiation of initially uniform clouds was studied by \cite{1994A&A...289..559L} and later by \cite{1996ApJ...458..222B}. \cite{1996ApJ...458..222B} considered the effects of the photodissociating as well as the ionizing radiation. They derived conditions under which the photodissociation front is able to overtake the ionization front and propagate into the shocked gas. If this occurs, the evolution of the cloud becomes rather more complicated. The conditions for the dissociation front to outrun the ionization front depend on the ratio of Lyman--continuum to far--UV photons emitted by the source (approximately unity, in the case of the source assumed here) and on the optical depth presented to the Lyman photons by the photoevaporation flow, resulting in a range of optical depths for which photodissociation should not be ignored. In our simulations, we find that the optical depth of the photoevaporation flow is $<0.1$, so that the photodissociation front cannot outrun the ionization front. Checking the validity of these assumptions is an important area of further work, but is beyond the scope of this paper. \cite{1994A&A...289..559L} divide the parameter space of target clouds into five regions. Our simulated cloud falls into their region V, in which an initially R--type front transitions to a D--type front and proceeds through the cloud. This is indeed what we observe.\\
\begin{figure*}
     \centering
     \subfigure[Column density map viewed along the $y$--axis of the bottom run. The ionizing source is marked by a green cross.]{\includegraphics[width=0.31\textwidth]{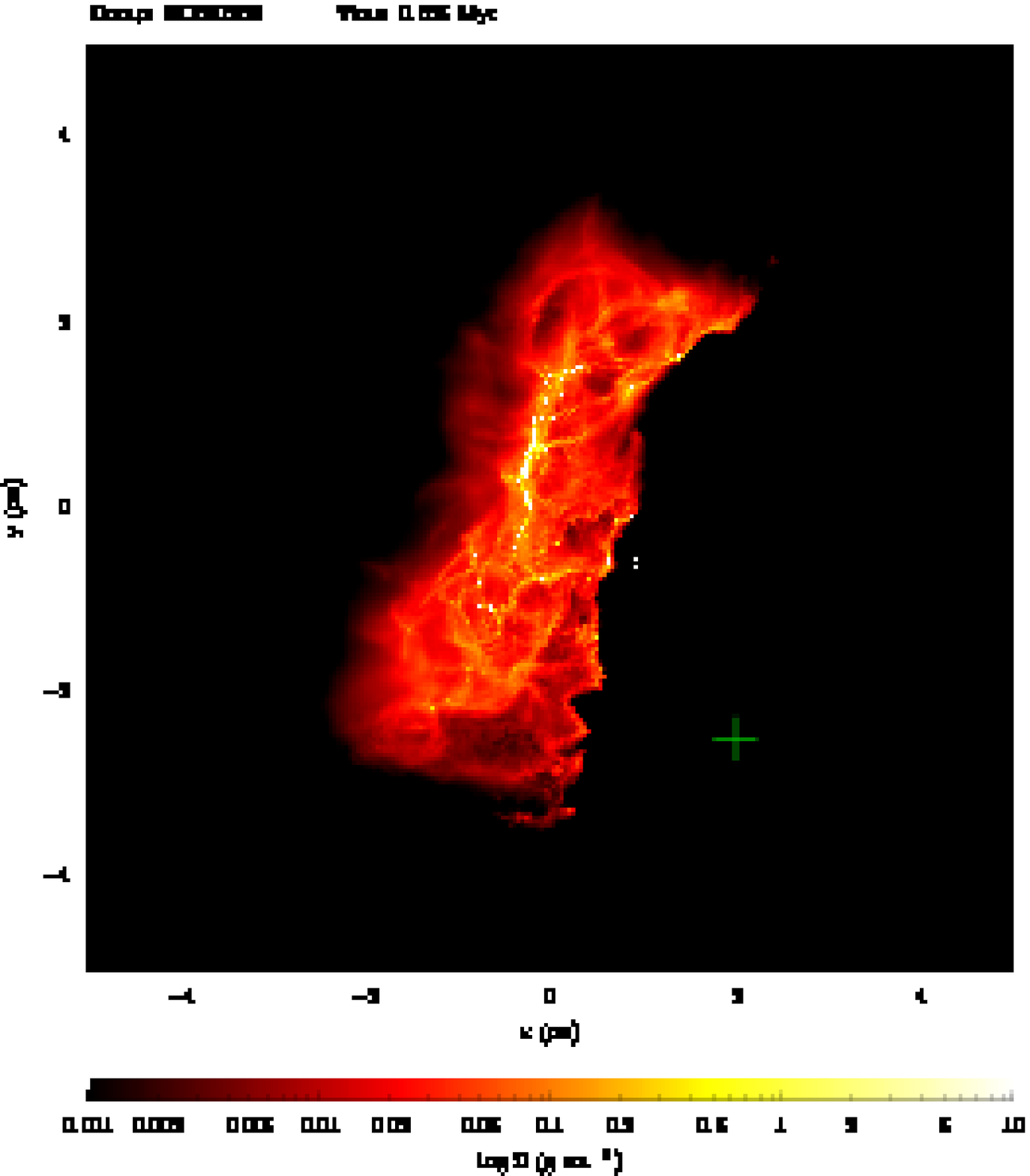}}     
     \hspace{.1in}
     \subfigure[Column density map viewed along the $y$--axis of the control run.]{\includegraphics[width=0.31\textwidth]{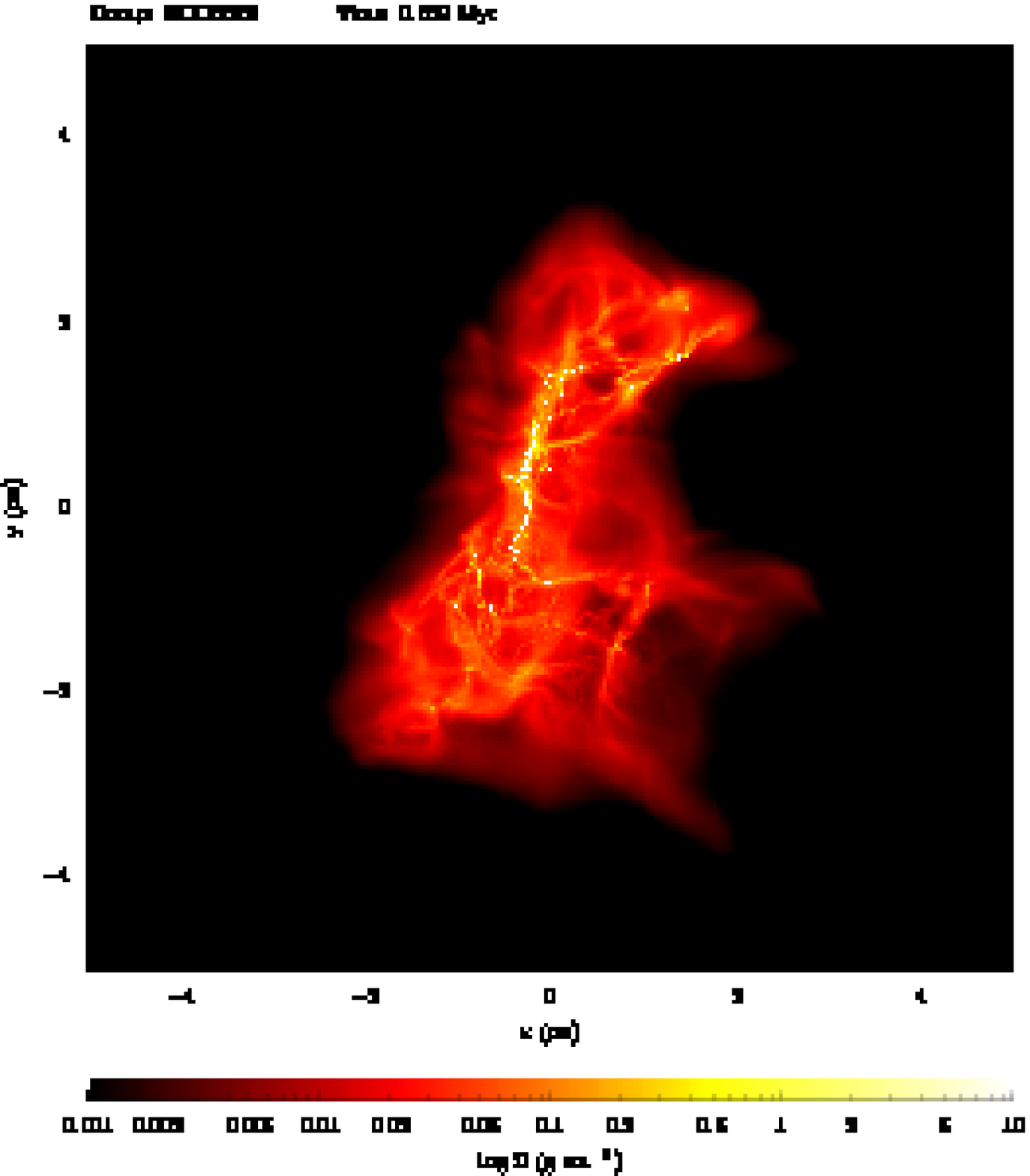}}
     \hspace{.1in}
     \subfigure[Column density map viewed along the $y$--axis of the top run at an age. The ionizing source is marked by a green cross.]{\includegraphics[width=0.31\textwidth]{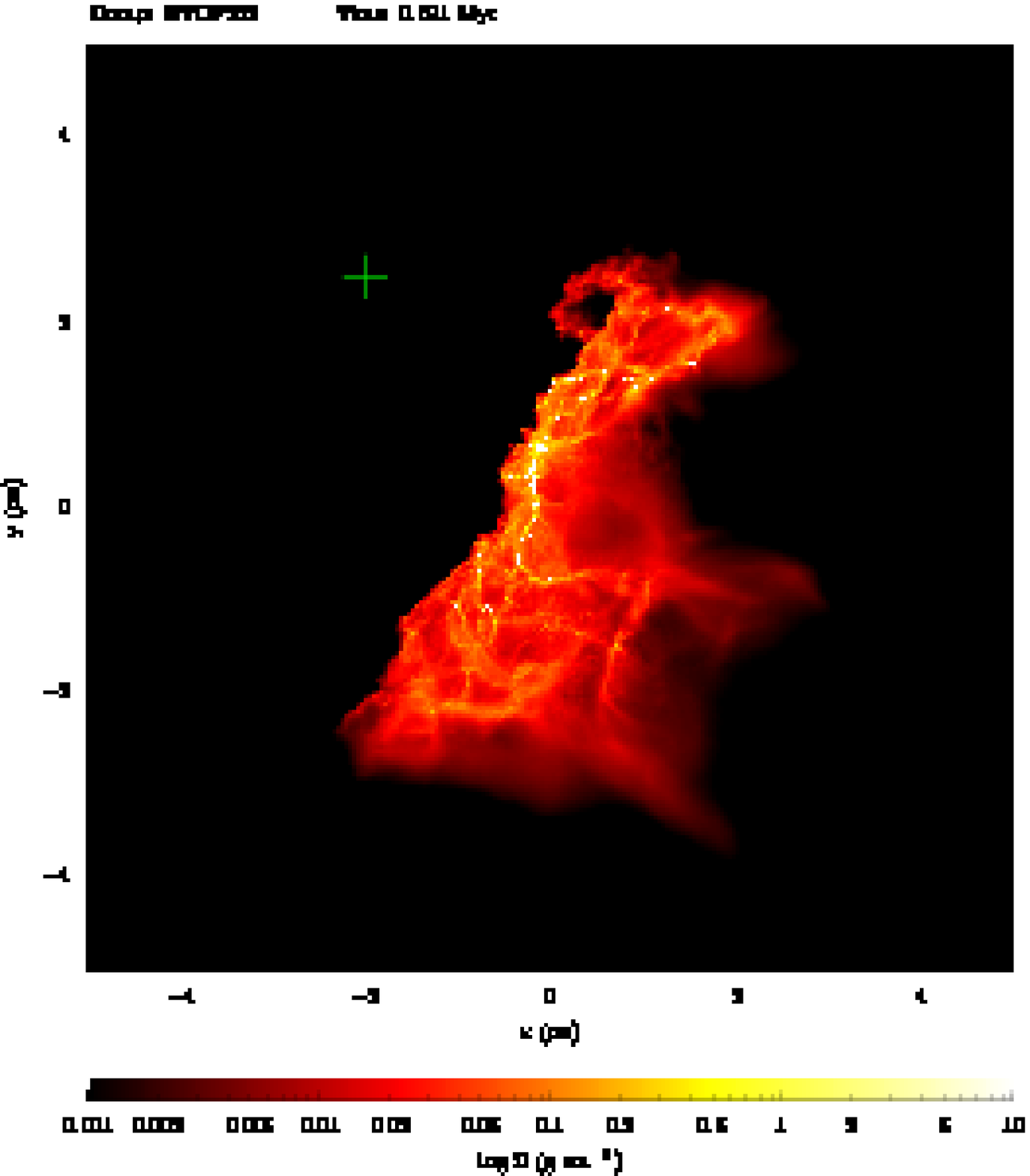}}
     \caption{Comparison of the morphologies of the clouds in the bottom (left panel), control (middle panel) and top (right panel) runs at $\sim0.67$ Myr. Actual times were chosen so that all runs have the same number of stars.}
   \label{fig:snaps}     
\end{figure*}   
\section{Initial conditions}
We construct a model of a molecular cloud of mass 2.3$\times10^{3}$ M$_{\odot}$. The cloud is modelled with $\sim1.65\times10^{6}$ particles, with a smooth gradient in particle mass of a factor of 3 in the $y$--direction, so that the positive--$y$ regions of the cloud are somewhat more bound than the negative--$y$ regions. The resulting minimum self--gravitating mass that can be resolved being $\approx0.05-0.15$M$_{\odot}$. The sinks have accretion radii of 0.002pc, corresponding to a density of $\sim3\times10^{-16}$g cm$^{-3}$, and we also smooth the sink--sink gravitational interactions at this scale. With this sink formation density, only the first segment of the equation of state given above is relevant and the sink particles formed should strictly be regarded as star--forming cores since their final stages are not followed, although we will continue to refer to them as `stars'. The cloud is ellipsoidal in shape, with the long axis being the $y$-axis and initial dimensions of $2\times2\times4$ pc. The density is initially uniform but the cloud is seeded with a Kolmogorov turbulent velocity field, so that it rapidly develops a complex density structure. The cloud is gravitationally bound overall, with a virial ratio of unity. We make three copies of the cloud. One, which we refer to as the `control' run is allowed to evolve undisturbed by ionizing radiation. A second (the `top' run) is illuminated from near its upper (more bound) end, with the ionizing source located at (-2.0, 2.5, 0.0)pc, while the third is illuminated from near the bottom (less bound) end from (2.0, -2.5, 0.0) pc. We will refer to the top and bottoms runs collectively as `feedback' runs. In both cases, the radiation source has an ionizing photon luminosity of $10^{49}$s$^{-1}$, roughly equivalent to a single $60$M$_{\odot}$ O--star or a small cluster of O--stars such as the Trapezium. The luminosity of the source and its proximity to the edge of the cloud (initially $\sim1$pc) results in a photon flux of $\sim8\times10^{10}$cm$^{-2}$s$^{-1}$. These parameters are intended to be realistic. The photon luminosity is equivalent to a lone high--mass O--star or to a low--mass (a few $\times10^{3}$M$_{\odot}$) cluster hosting several lower--mass O--stars. A radius or order 1pc is representative for a cluster of such a size \citep[e.g][]{2009ApJ...691..946M}. If, in the more likely scenario, the ionizing source is regarded as being a small stellar cluster, it cannot therefore be placed much closer to the model cloud (particularly as we are ignoring any gravitational interaction between the source of radiation and the clouds). We therefore consider our chosen flux to be on the high side of what is astrophysically realistic given the size and mass of our target cloud. \cite{2010arXiv1007.2727B} consider the problem of the irradiation of clouds modelled as Bonner--Ebert spheres by ionizing sources of various luminosities sited 3pc from the initial cloud edge. Our chosen flux lies in the middle of the range of fluxes considered by these authors and inside the range where the ionization triggers star formation, rather than destroying their model clouds.\\
\section{Results}
\subsection{Cloud morphology}
In our previous calculations \citep{2007MNRAS.377..535D}, we found that ionizing radiation had a profound effect on the cloud's morphology during the $\sim0.5$ freefall times for which we ran the simulation. In this paper, we consider lower--mass clouds of smaller spatial extent and higher density. We find that the effect on the cloud morphology of the irradiation is less pronounced because the gas in our new calculations is denser, so that the ionization initially evaporates less material before the ionization front transitions from R--type to D--type, and erosion of the cloud proceeds more slowly in comparison to the cloud freefall time, which is shorter in our new simulations.\\  
\indent In Figure \ref{fig:snaps}, we show column--density plots of the bottom, control and top runs at comparable times. We terminated the simulations after $\sim0.7$Myr because the most massive stars in all runs are approaching 10M$_{\odot}$ by this time and hence are almost massive enough to be ionizing sources themselves. We did not wish to complicate our simulations by attempting to model the effects of external and internal feedback, so we stop at this point. The colours in Figure \ref{fig:snaps} denote gas column density and white dots are sink particles. Ionizing sources are marked by green crosses where appropriate. In the control run, the gas is distributed in a network of dense filaments, and it is here that most of the star formation takes place, with the majority being in the largest filament running down most of the length of the cloud. In the bottom run, there is a considerable quantity of molecular gas between the source and the main filament, which partially prevents the feedback from strongly influencing the evolution of the gas in this region. Instead, the observable effects of photoionization in this run are largely to compress the bottom right portion of the cloud and drive the material there towards the cloud core. Conversely, in the top run the ionizing radiation impinges almost directly on the main star--forming filament and smears it out to some extent along an axis pointing away from the source.\\
\begin{figure}
\includegraphics[width=0.45\textwidth]{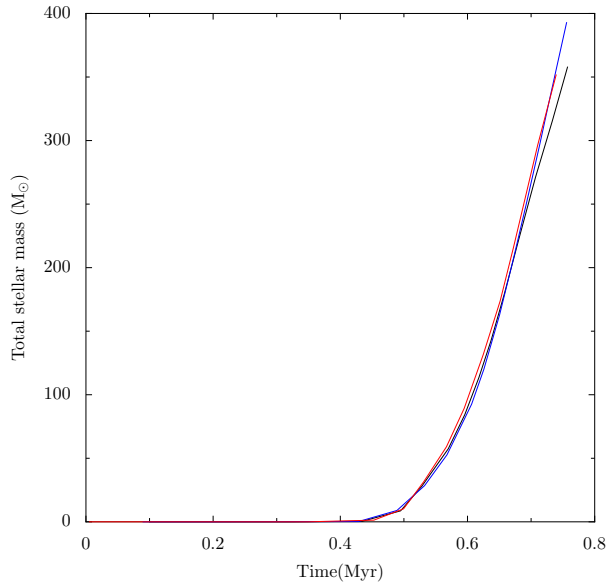}
\caption{Total mass in stars as a function of time in the control run (back line), the top run (blue line) and the bottom run (red line).}
\label{fig:msinkt}
\end{figure}
\begin{figure}
\includegraphics[width=0.45\textwidth]{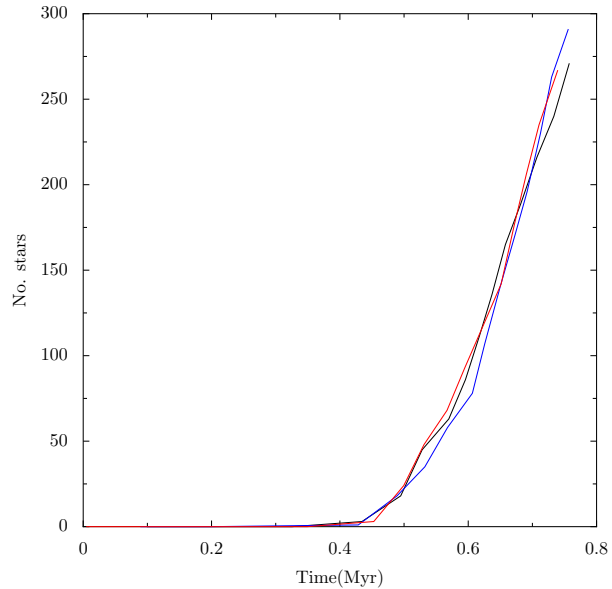}
\caption{Total number of stars as a function of time in the control run (black line), the top run (blue line) and the bottom run (red line).}
\label{fig:nsinkt}
\end{figure}
\begin{figure}
\includegraphics[width=0.45\textwidth]{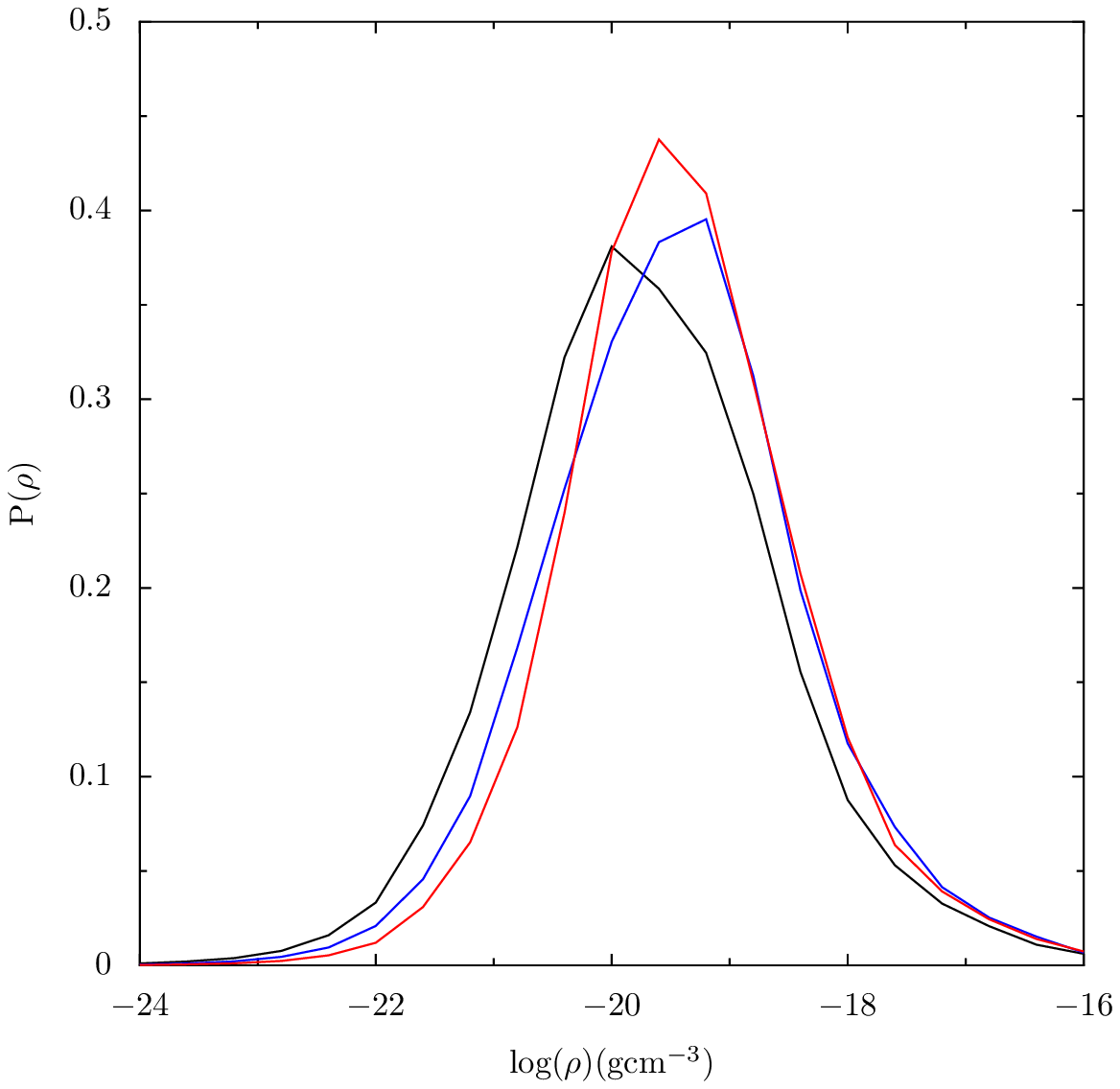}
\caption{Probability density functions for the gas density in the control run (black line), the top run (blue line) and the bottom run (red line).}
\label{fig:dens}
\end{figure}
\begin{figure}
\includegraphics[width=0.45\textwidth]{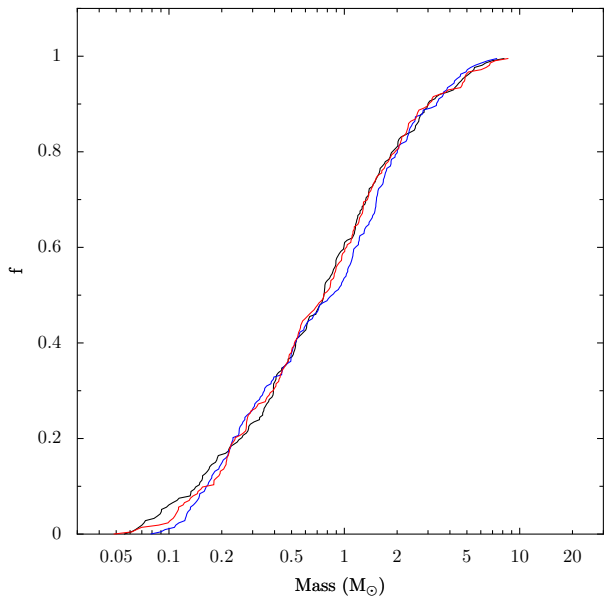}
\caption{Cumulative mass functions of the control run (black line), the top run (blue line) and the bottom run (red line) at times when each run contains 213 stars.}
\label{fig:cum_mf}
\end{figure}
\subsection{Star formation rate and efficiency}
In Figures \ref{fig:msinkt} and \ref{fig:nsinkt}, we plot the total stellar mass and the total numbers of stars as functions of time in the three runs. These figures show that external irradiation has little effect on either the total quantity of mass involved in star formation (i.e. the star formation efficiency) or the total numbers of stars formed. In Figure \ref{fig:dens}, we plot the probability density function of the gas density in the control run and the feedback runs. The functions are shifted to higher densities in the feedback runs, but not by a large factor, indicating that the external irradiation has a rather modest effect on the density structure of the cloud, and hence has little influence on the rate of star formation.
\subsection{Stellar mass functions}
In comparing the mass functions of the three simulations, it is the shape of the function that concerns us. Since the simulations form stars at different rates, comparing them at given times is misleading, since then the mass functions contain different numbers of objects. In Figure \ref{fig:cum_mf}, we plot cumulative mass functions from all three simulations at times chosen so that all contain 213 stars. We see that the mass functions are very similar, although the bottom--triggered run shows a deficit of objects around 1 M$_{\odot}$, and is overall somewhat steeper. We perform Kolmogorov--Smirnov tests on these mass functions to determine the probability that they have the same mass function. We find that the probabilities that the control and top--triggered mass functions are drawn from the same distribution is $18\%$, that the control and bottom--triggered are is $74\%$ and that the two triggered mass functions are drawn from the same distribution is $24\%$. These differences are not statistically significant. \\
\subsection{Triggered stars}
\begin{figure*}
     \centering
     \subfigure{\includegraphics[width=0.45\textwidth]{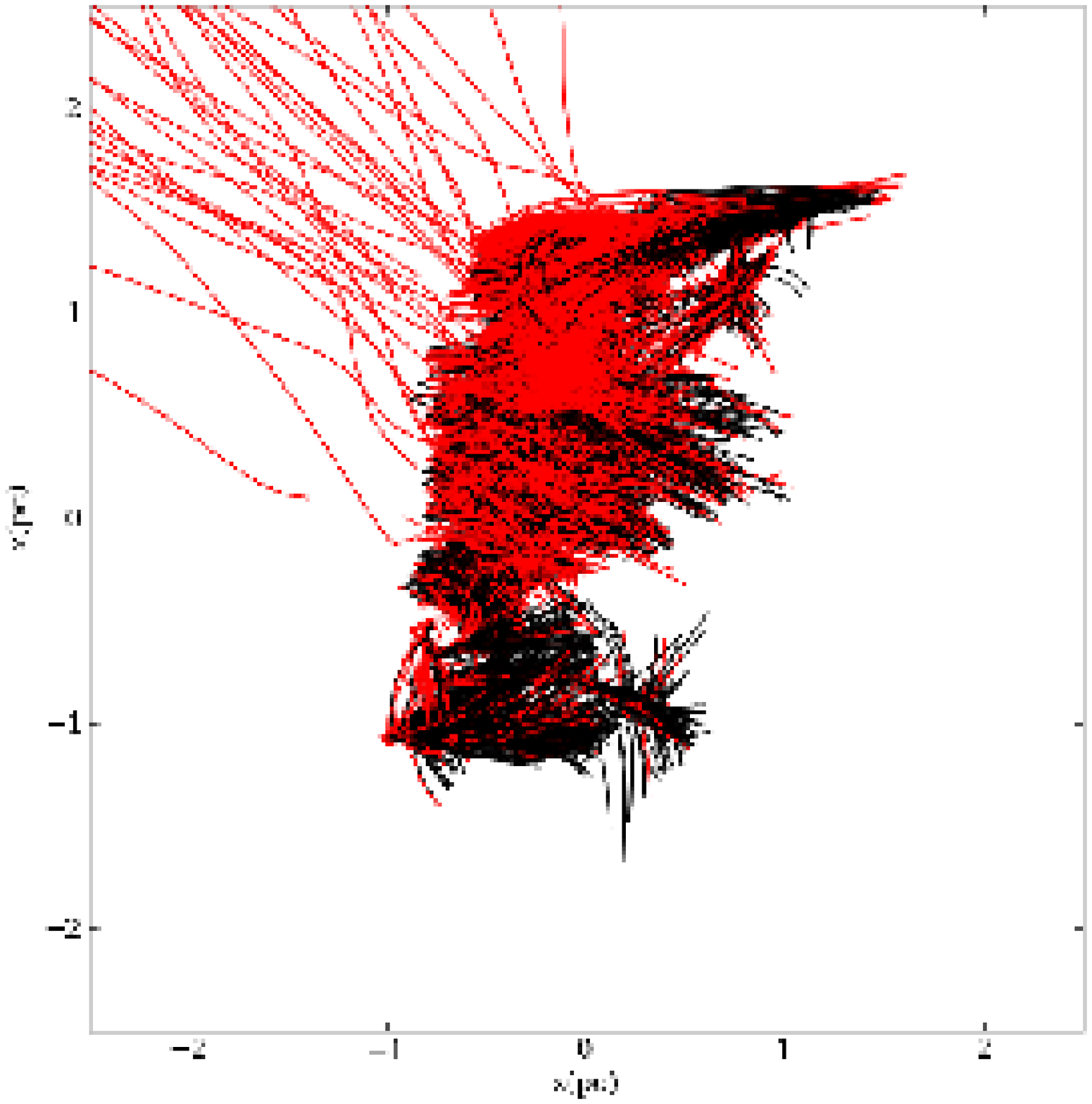}} 
     \hspace{.1in}
     \subfigure{\includegraphics[width=0.45\textwidth]{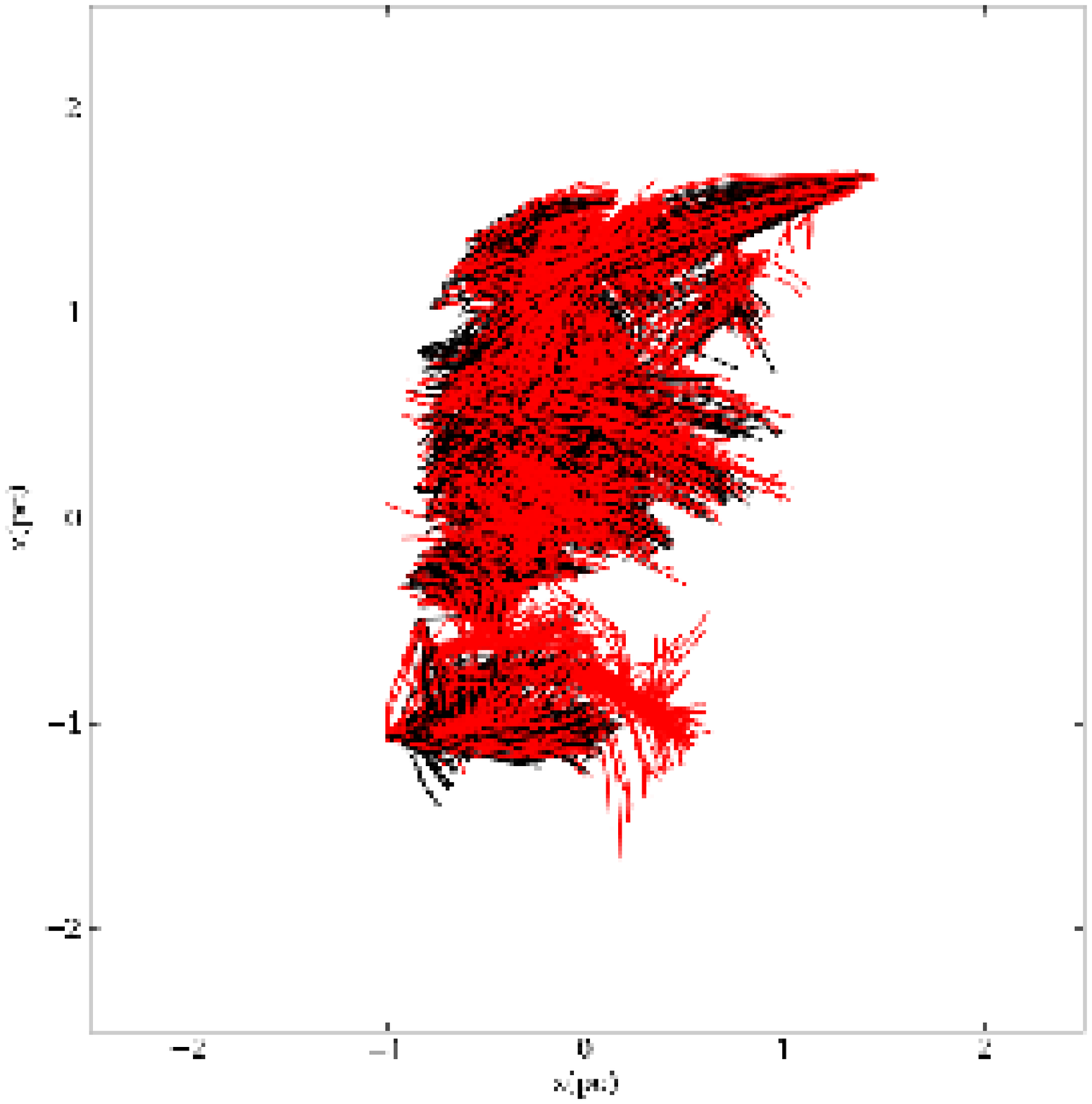}}
     \caption{Traces in the top--triggered (left panel) and bottom--triggered (right panel) simulations of gas particles involved in star formation in the control run. Particles traced in black are those that are also involved in star formation in each ionized run, whereas those marked in red are involved in star formation only in the control run. For clarity, only one particle in ten is plotted.}
     \label{fig:trace_control}
\end{figure*}

\begin{figure*}
     \centering
     \subfigure{\includegraphics[width=0.45\textwidth]{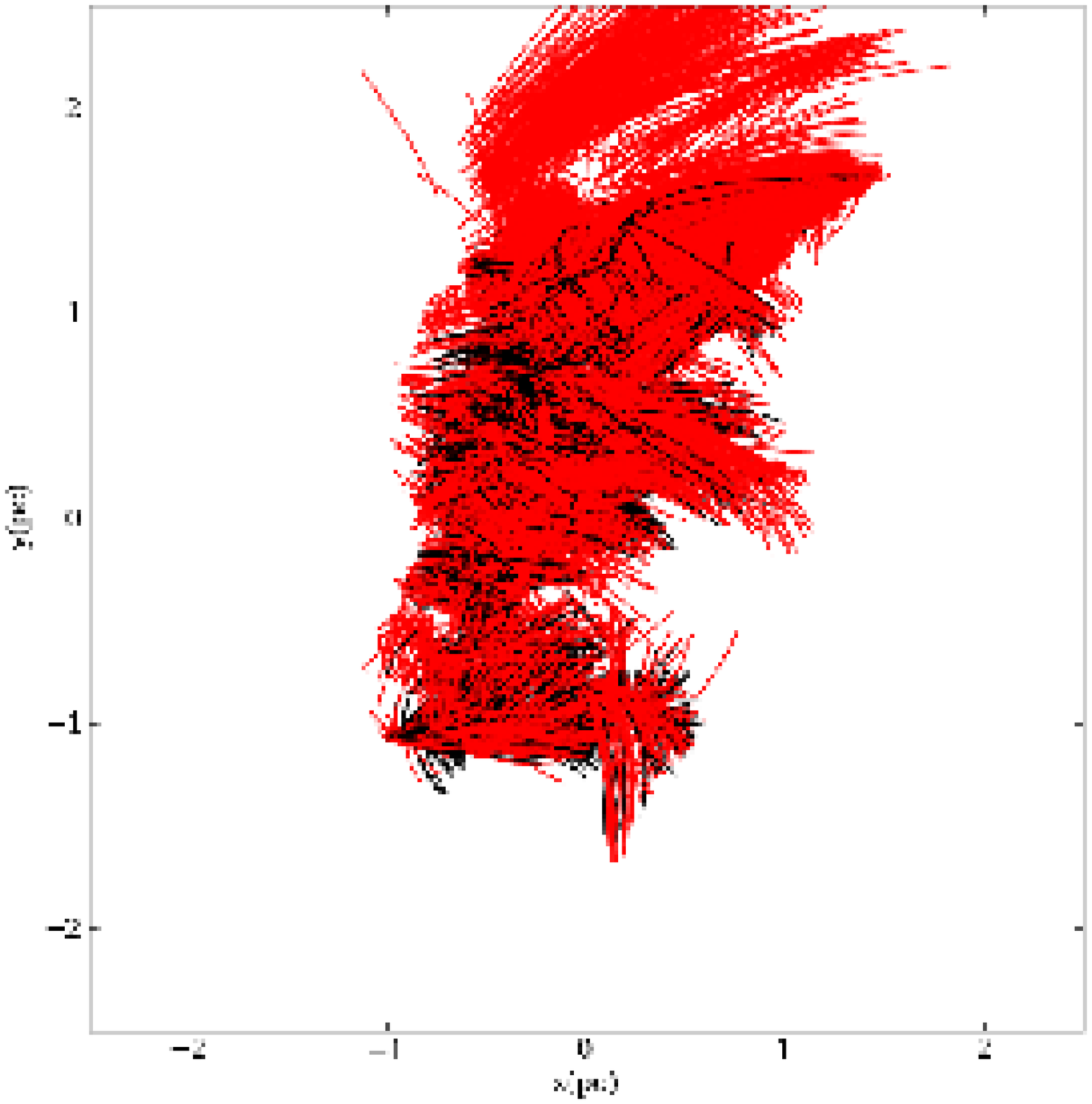}} 
     \hspace{.1in}
     \subfigure{\includegraphics[width=0.45\textwidth]{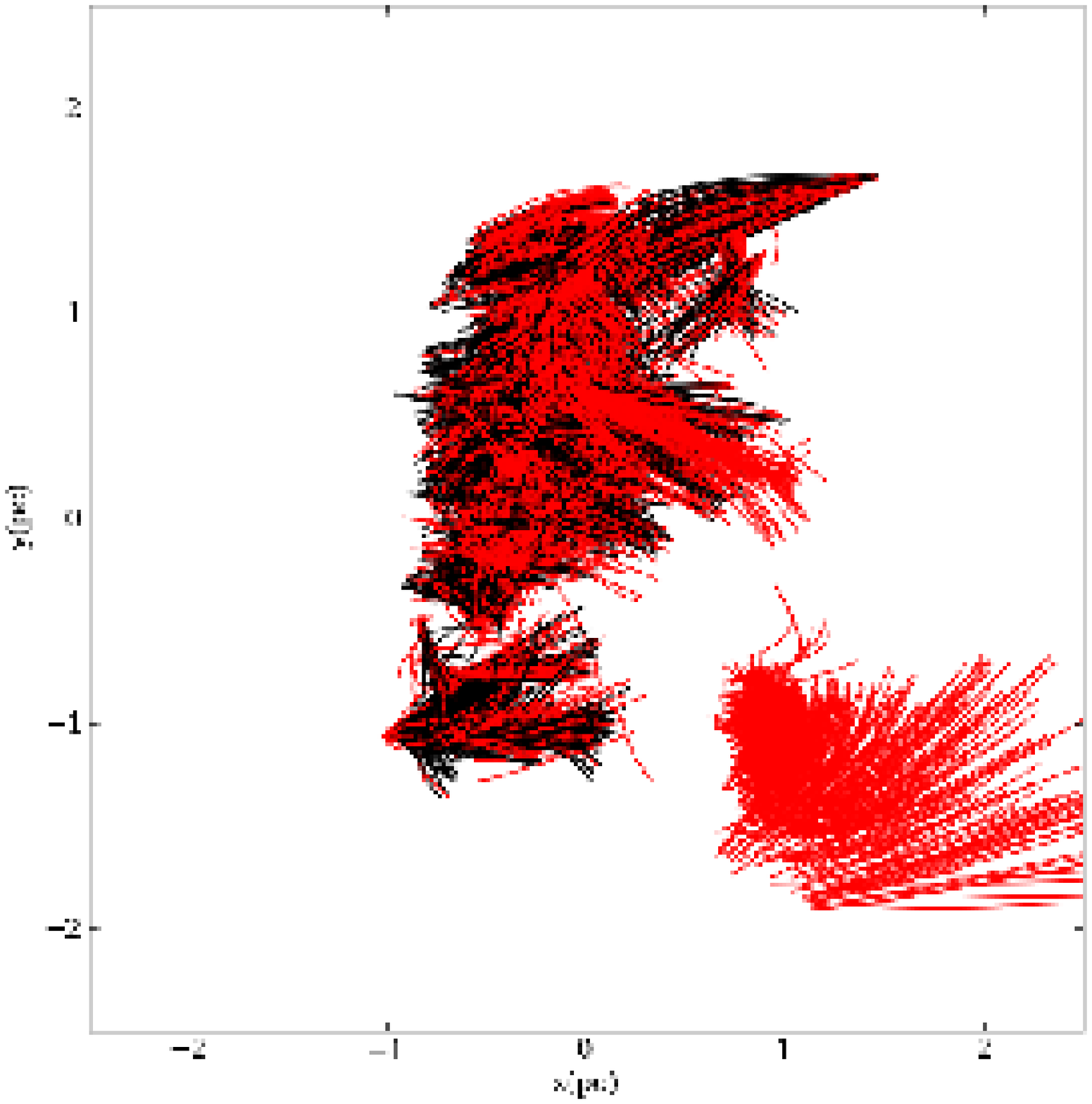}}
     \caption{Traces in the control run of gas particles involved in star formation in the top--triggered (left panel) and bottom--triggered (right panel) simulations. Particles traced in black are those that are also involved in star formation in the control run, whereas those marked in red are involved in star formation only in each ionized run. For clarity, only one particle in ten is plotted.}
     \label{fig:trace_bound}
\end{figure*}

\begin{figure}
     \centering
     \subfigure{\includegraphics[width=0.23\textwidth]{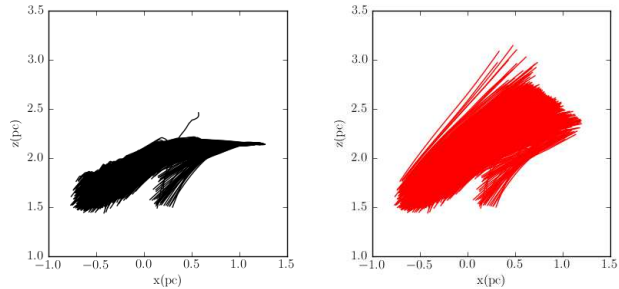}} 
     \hspace{.01in}
     \subfigure{\includegraphics[width=0.23\textwidth]{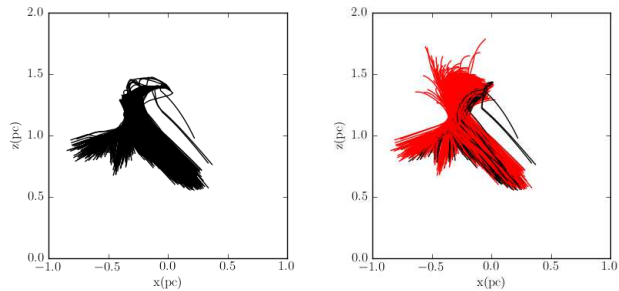}}
     \caption{Traces of gas particles involved in the formation in the top--triggered run of a sink particle (left panel) and of the same gas particles in the control run (right panel). Particles traced in black are those that are involved in star formation in the control run, whereas those marked in red are involved in star formation only in the top--triggered run.}
     \label{fig:trig_sink1}
\end{figure}

\begin{figure}
     \centering
     \subfigure{\includegraphics[width=0.23\textwidth]{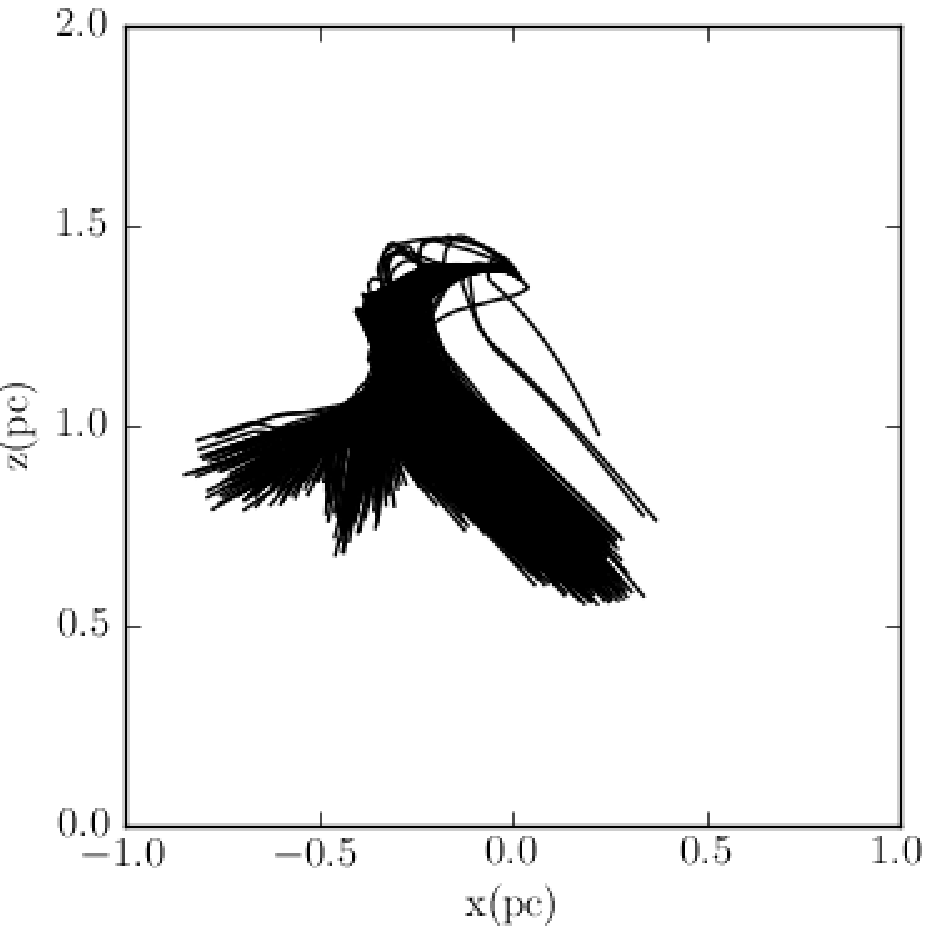}} 
     \hspace{.01in}
     \subfigure{\includegraphics[width=0.23\textwidth]{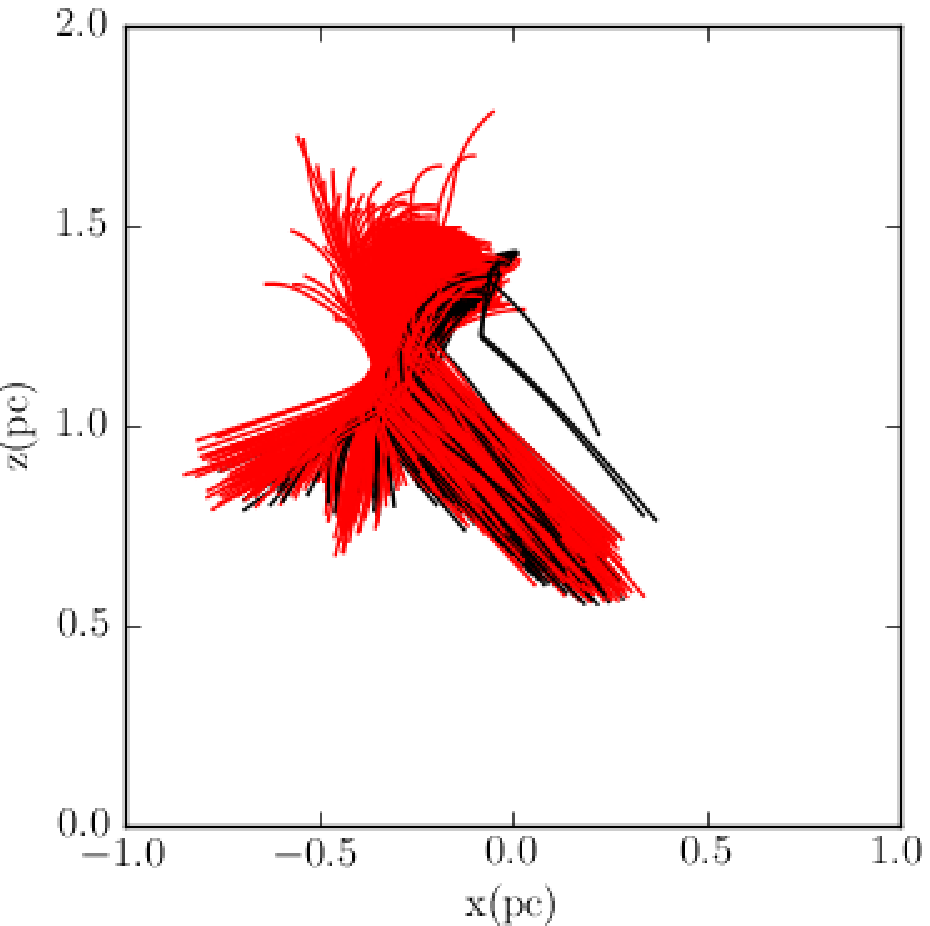}}
     \caption{Traces of gas particles involved in the formation in the top--triggered run of a sink particle (left panel) and of the same gas particles in the control run (right panel). Particles traced in black are those that are involved in star formation in the control run, whereas those marked in red are involved in star formation only in the top--triggered run.}
     \label{fig:trig_sink2}
\end{figure}

\begin{figure*}
     \centering
     \subfigure{\includegraphics[width=0.33\textwidth]{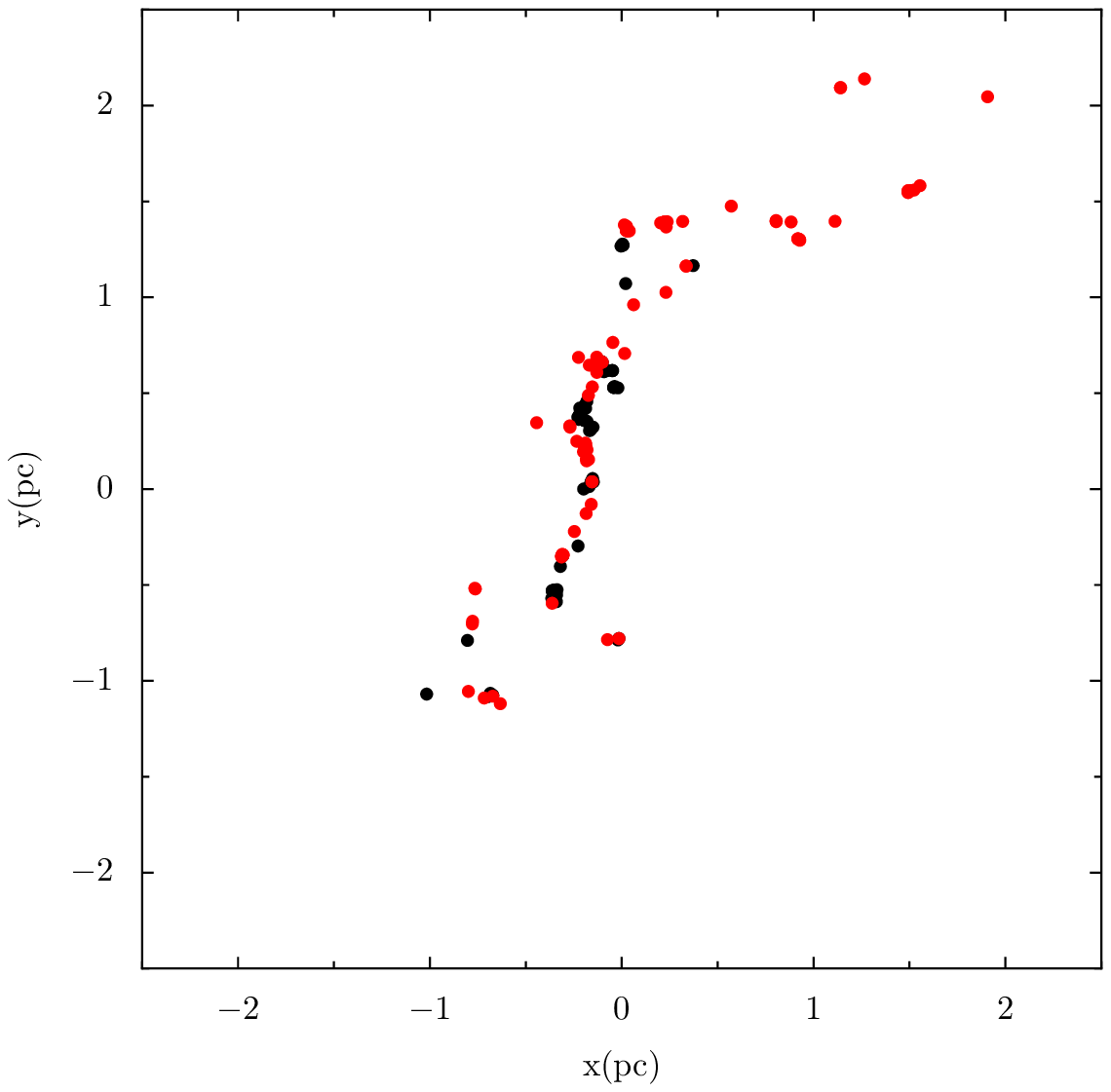}}
     \subfigure{\includegraphics[width=0.33\textwidth]{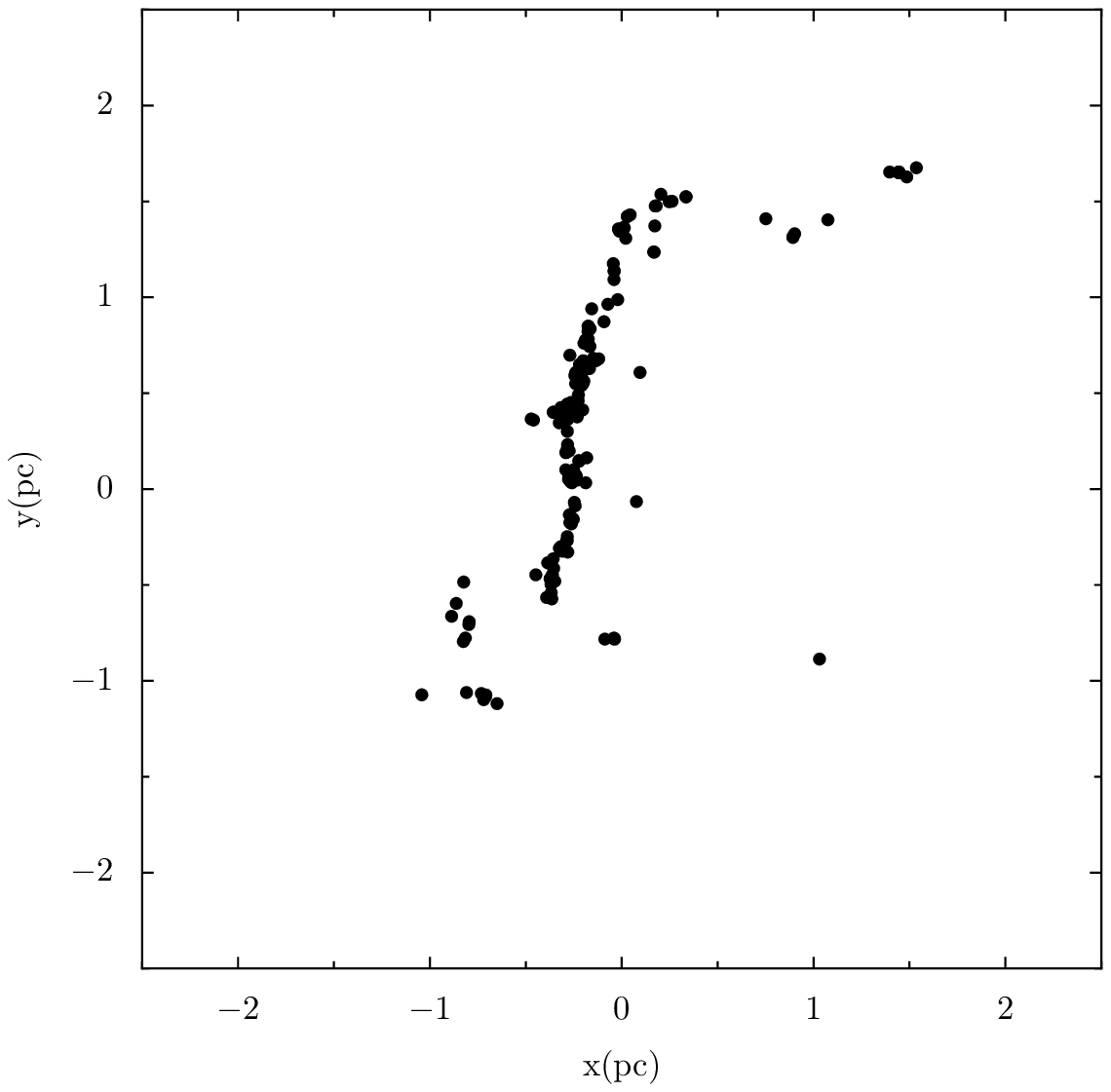}}    
     \subfigure{\includegraphics[width=0.33\textwidth]{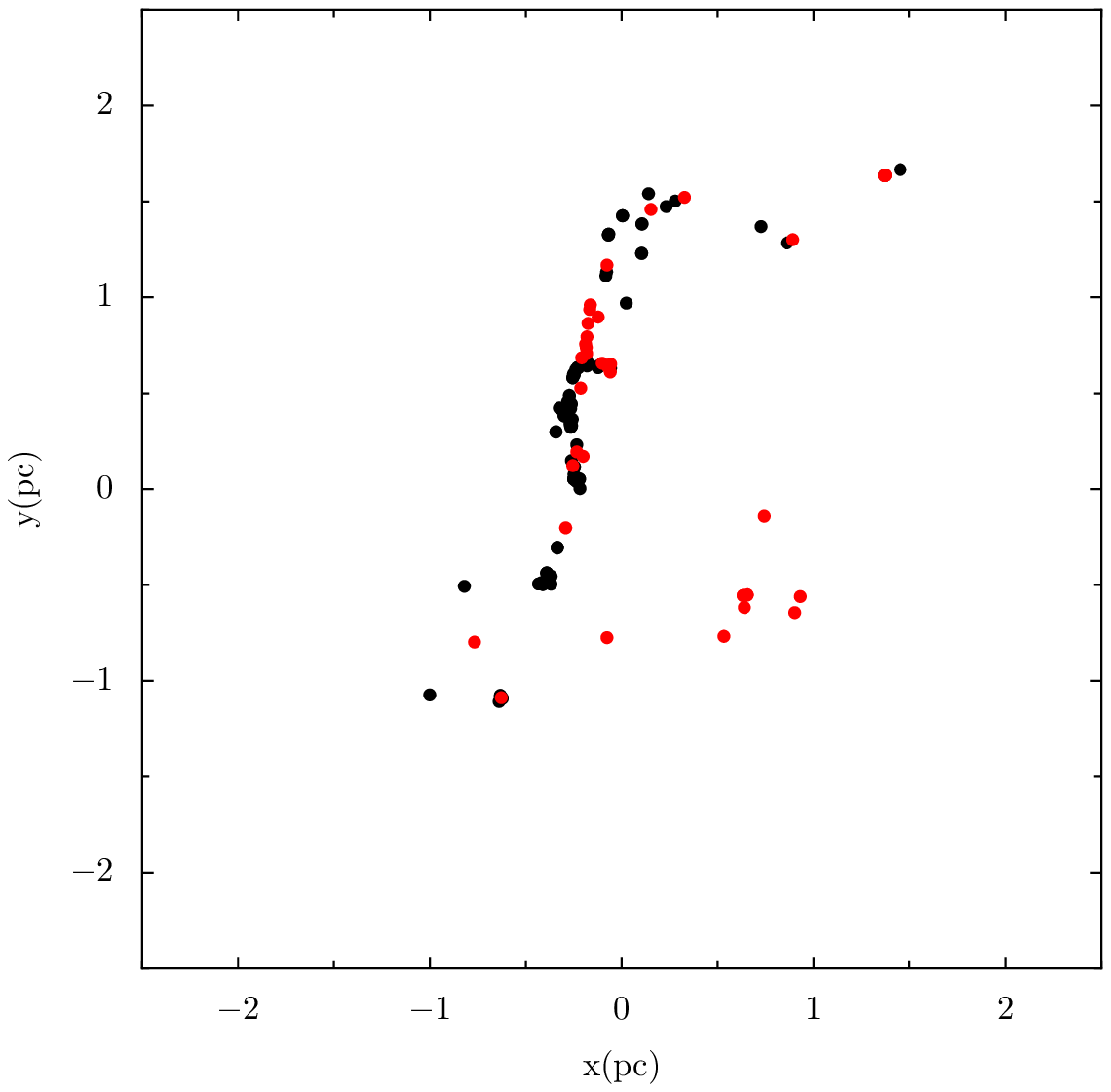}}
     \caption{Locations of triggered (red) and spontaneous (black) stars according to the criterion given in the text in the top--triggered (left panel) and bottom--triggered (right panel) simulations, with all stars in the control run shown in the centre panel for comparison. The top run contains 103 triggered stars and 93 untriggered stars (total, 196 stars), while the bottom run contains 39 triggered stars and 167 untriggered stars (total, 206 stars). The times of the plots are, respectively, 0.69, 0.66 and 0.66 Myr in each run.}
       \label{fig:trig} 
\end{figure*}

\begin{figure*}
     \centering
     \subfigure{\includegraphics[width=0.31\textwidth]{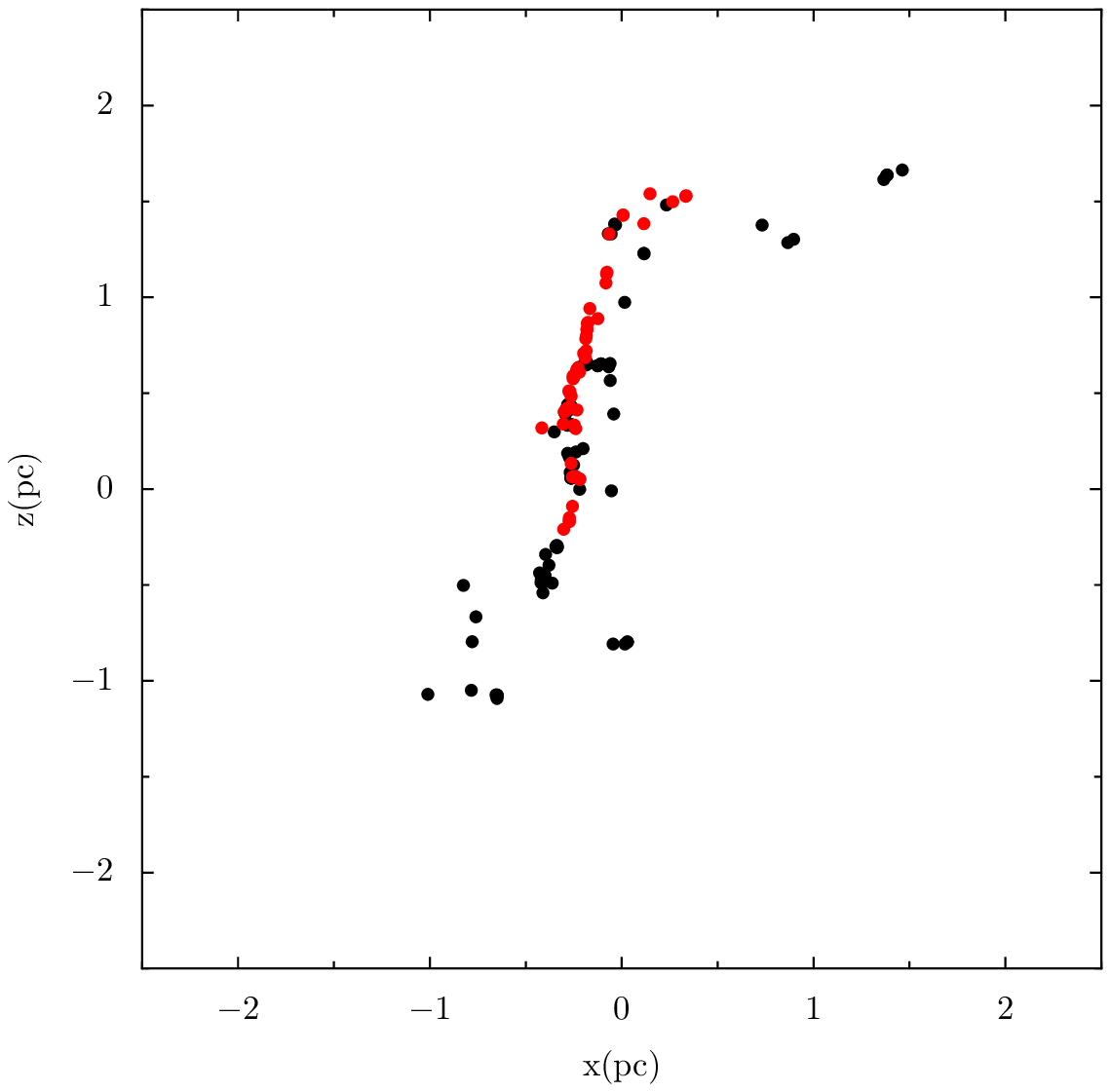}} 
     \hspace{.1in}
     \subfigure{\includegraphics[width=0.31\textwidth]{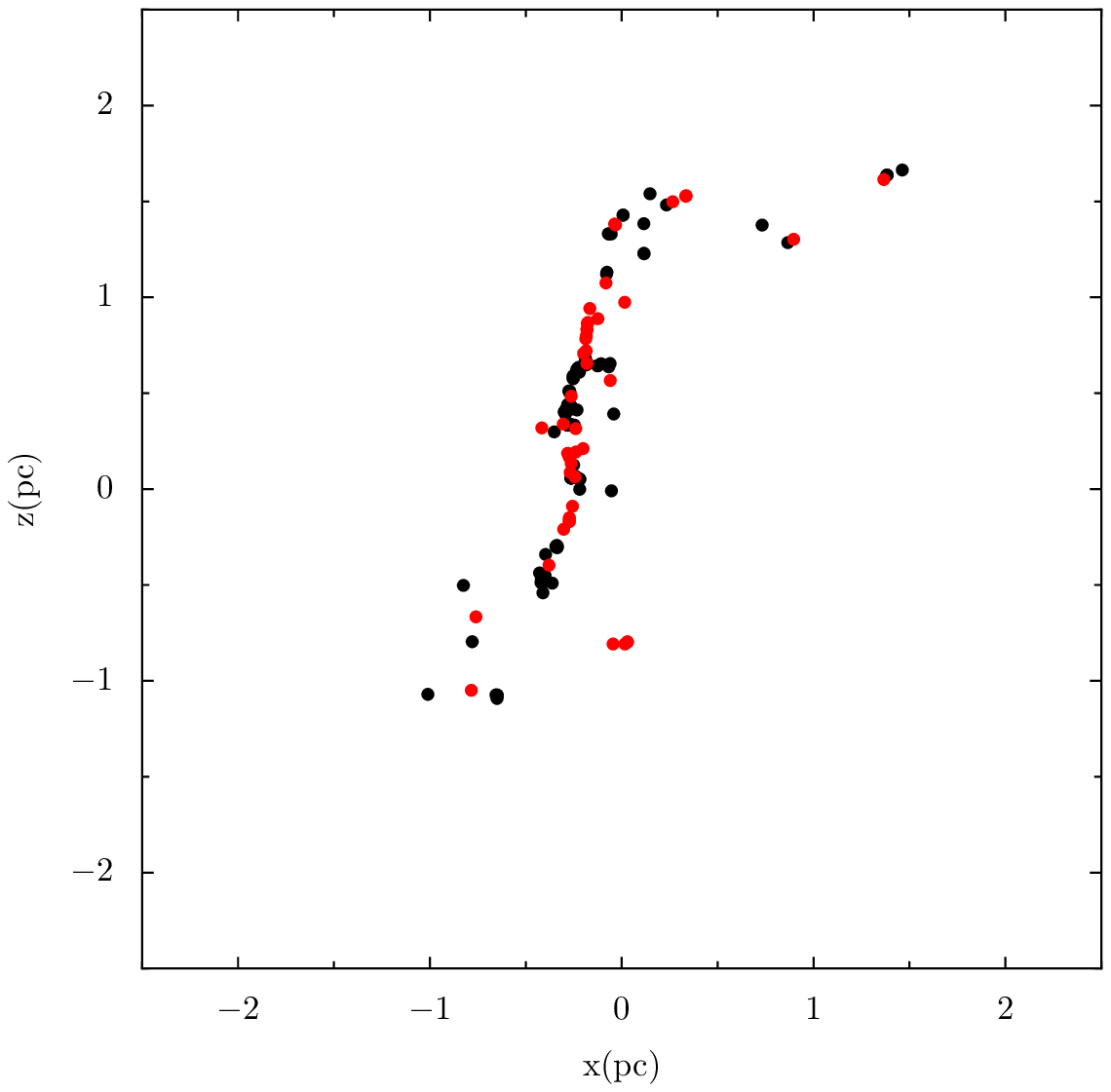}}
     \caption{Locations of stars in the control run which do form (black) or are aborted (red) in the top--triggered (left panel) and bottom--triggered (right panel) simulations, at a time of 0.66Myr in the control run.}
     \label{fig:abort}
\end{figure*}

\begin{figure*}
     \centering
     \subfigure[Column density map viewed along the $z$--axis of the top right corner of the top run at 0.499 Myr.]{\includegraphics[width=0.31\textwidth]{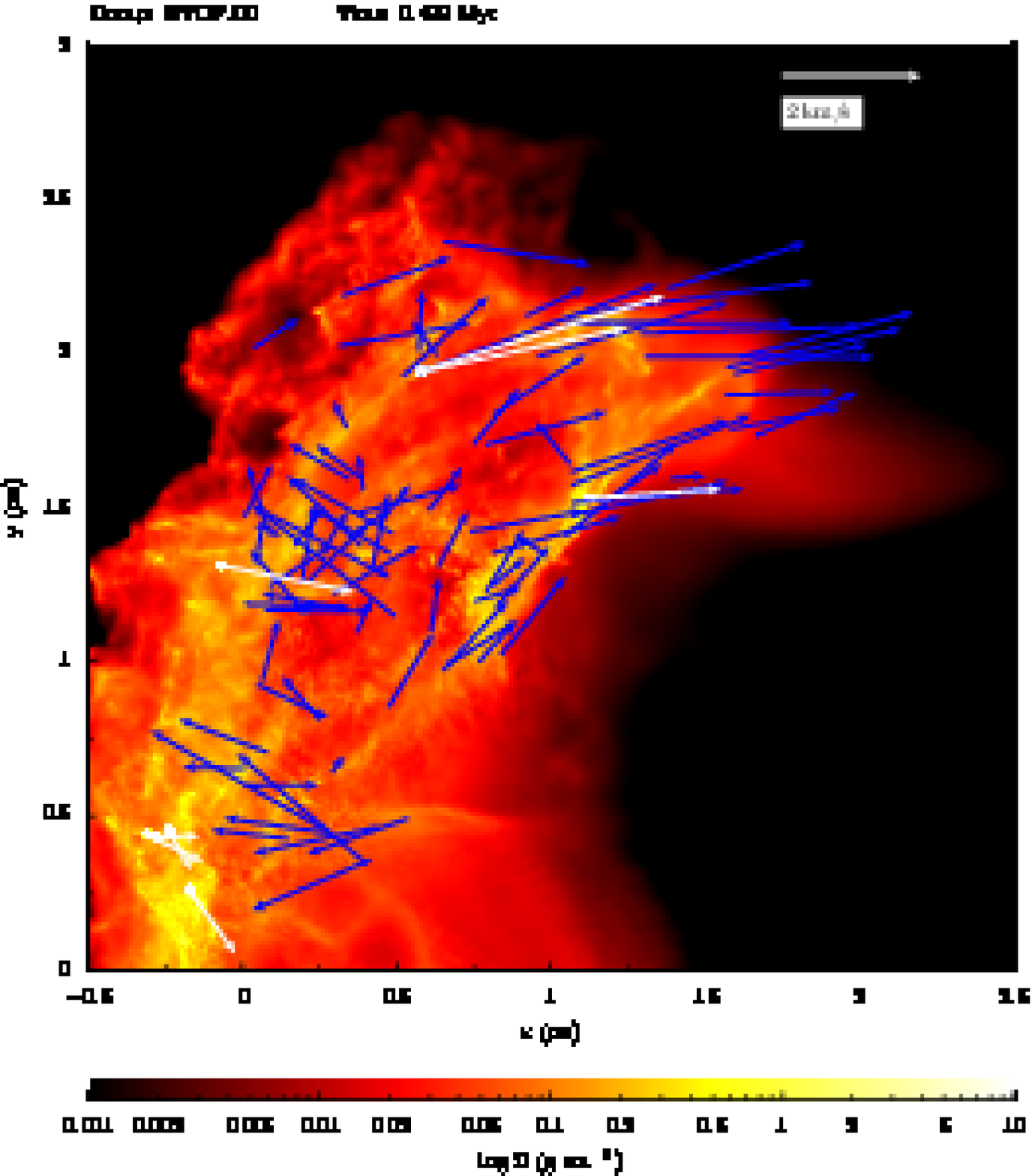}}     
     \hspace{.1in}
     \subfigure[Column density map viewed along the $z$--axis of the top right corner of the top run at 0.578 Myr.]{\includegraphics[width=0.31\textwidth]{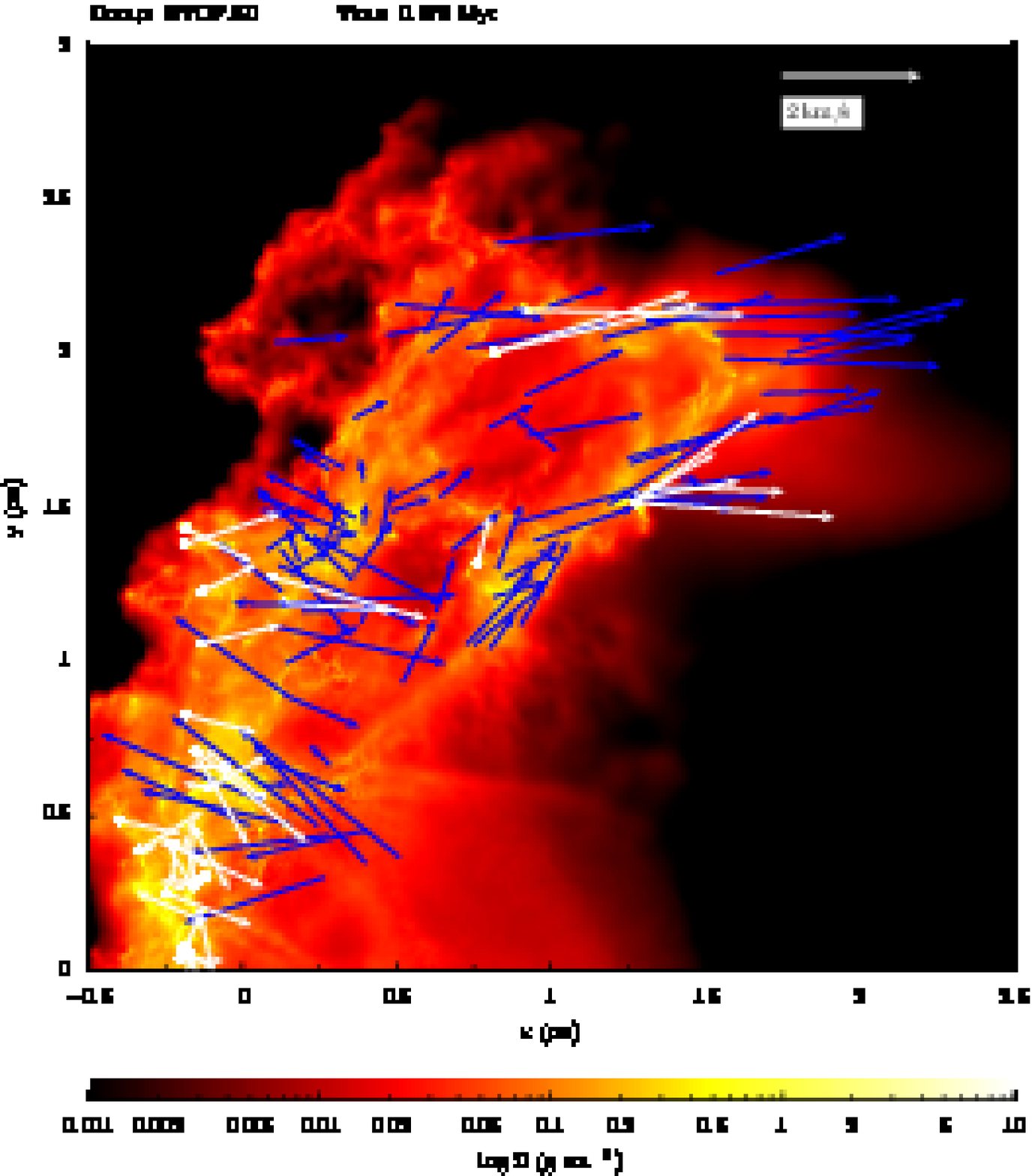}}
     \hspace{.1in}
     \subfigure[Column density map viewed along the $z$--axis of the top right corner of the top run at 0.636 Myr.]{\includegraphics[width=0.31\textwidth]{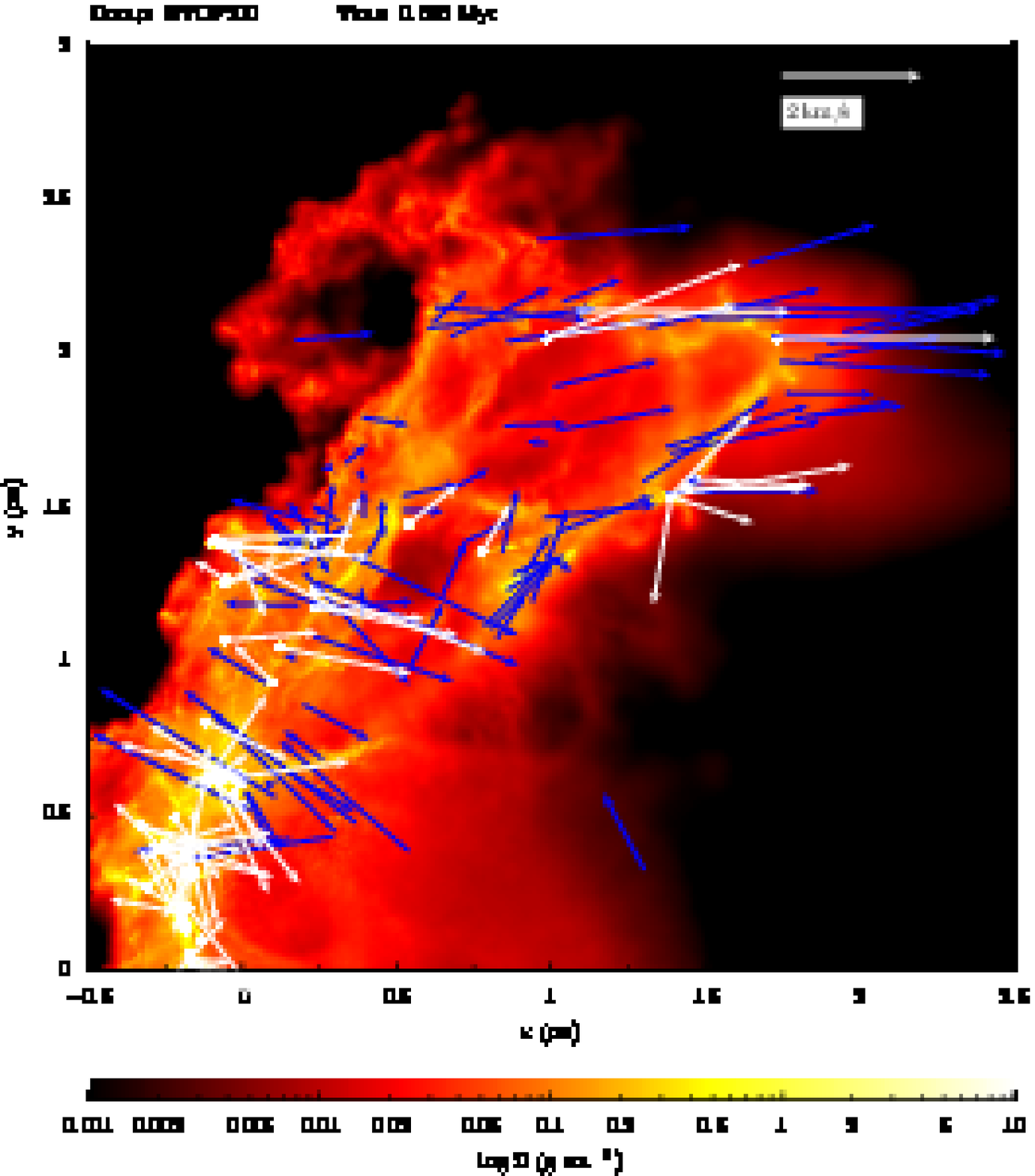}}
     \caption{Motions of the stars (white arrows) and randomly--selected gas particles (blue arrows) in the top right corner of the top run at three different epochs.}
        \label{fig:trigtopright}
\end{figure*}

\begin{figure}
\includegraphics[width=0.45\textwidth]{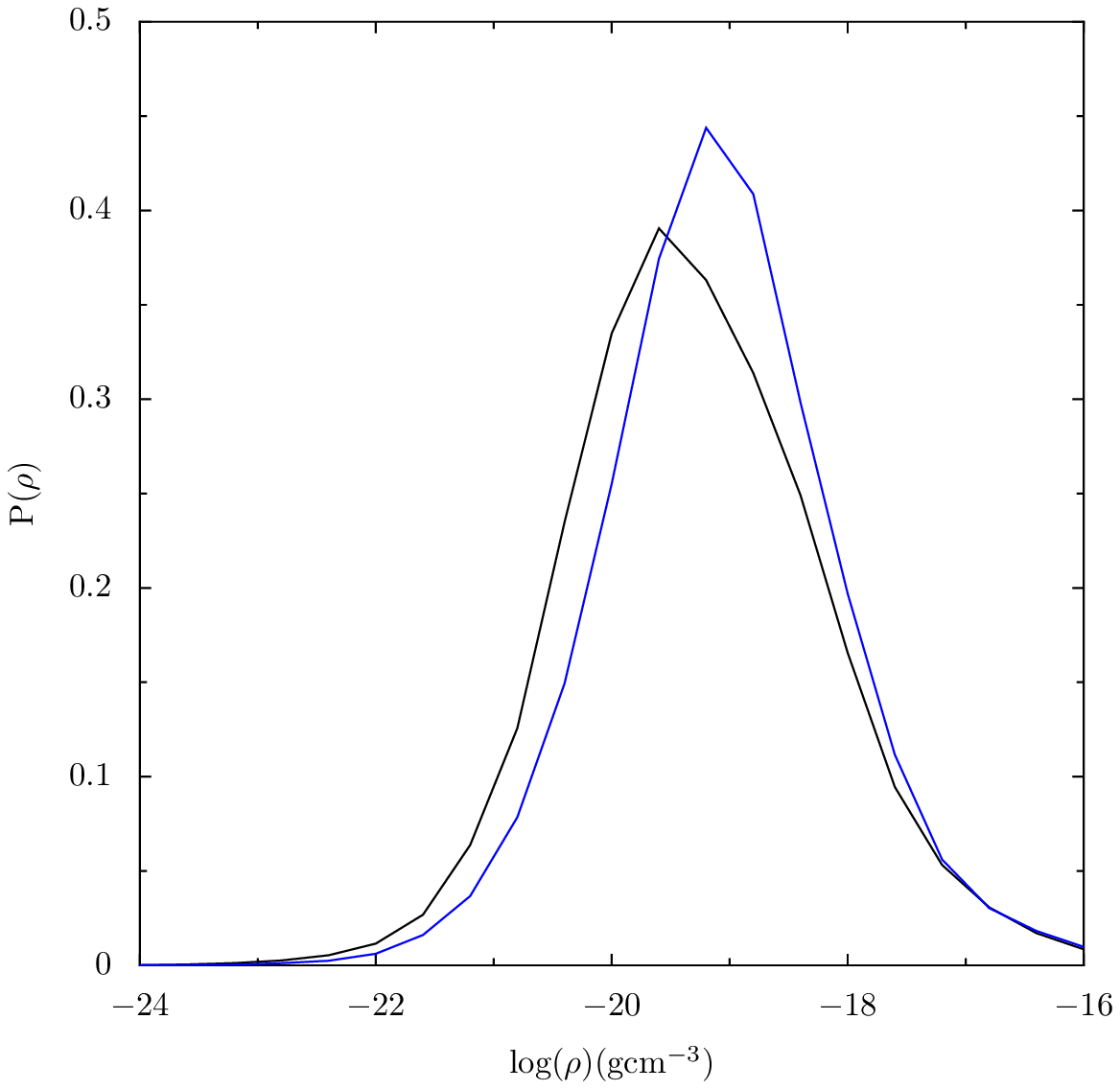}
\caption{Comparison of the density PDFs of the top run (blue) and the control run (black) in the region shown in Figure \ref{fig:trigtopright} at a time of 0.636Myr.}
\label{fig:dcomp_top}
\end{figure}

In \citep{2007MNRAS.377..535D} we defined strong triggering to mean the formation through the action of feedback of a star or star--forming core which failed to form in a control run evolving in the absence feedback from the same initial conditions. Since SPH is a Lagrangian technique, stars or cores can be unambiguously identified between simulations by the gas particles from they formed. Conversely, abortion was defined as the formation in the control run of a star/core which failed to form in the feedback run. We traced the histories of the $\sim100$ seed gas particles from which sinks initially formed in the absence of feedback and which were induced to form or prevented from forming in the feedback run. This technique, which we call the `same--seed' method, determines whether the \emph{same star formation event} occurs in two simulations starting from the same initial conditions and worked well in our previous calculations, where star formation occurred in a small number of isolated regions within a large globally--unbound cloud. However, star formation in the simulations presented in this work is much more vigorous, since the clouds are bound, and occurs largely in the same crowded region of all three copies of the parent cloud. Applying the same--seed method revealed that it was exceedingly rare (at the $\sim$1\% level) for the same $\sim$ 100 particles to form a seed in more than one simulation. Since very few sink--seeds in any run have counterparts in another run, almost all stars in the feedback runs would be triggered, and almost all in the control run aborted by this definition. Since the numbers, mass function and spatial distribution of stars in all three calculations presented here are very similar, this result is highly counterintuitive.\\
\indent We improved upon the same--seed method by tracing instead \emph{all} the particles going to form a given sink -- the seed particles from which the sink formed \emph{and} the particles accreted afterwards. If more than half of the particles forming a sink in one run also form a sink in another run, the \emph{same star} can be said to form in both calculations and we refer to this as the `same--star' method. Applying this criterion here, we find that the same star forms in two or more runs on only $\sim$20\% of occasions, even though the general morphology of star formation is very similar in all three calculations. This technique would also label most stars as triggered or aborted. The reason for these two results is that in all calculations most of the stars form in a small fraction of the cloud's volume. Even small perturbations to the gas density and velocity fields due to ionization can therefore affect exactly which gas particles are involved in forming a given object, making it unlikely that the same seed or the same star will form more than once.\\
\indent We illustrate this point in Figure \ref{fig:trace_control} where we show traces (i.e. the historical tracks) in the ionized runs of gas particles involved in star formation in the control run. Particles which are also involved in star formation in each ionized run are plotted in black, whereas particles only involved in star formation in the control run are plotted in red (only one tenth of the particles are plotted for reasons of clarity). The counterpart to Figure \ref{fig:trace_control} is Figure \ref{fig:trace_bound} where we plot the traces in the control run of particles from which stars form in the ionized runs. Particles which are also involved in star formation in the control run are plotted in black whereas those which are involved in star formation only in each ionized run are plotted in red. In Figure \ref{fig:trace_bound}, regions of the clouds close to the ionizing sources (the top right in the top--triggered run and the bottom left in the bottom--triggered run) contain large quantities of gas which only forms stars in the feedback runs. In the left panel of Figure \ref{fig:trace_control} by contrast, we see that some of the gas involved in star formation in the control run is actually ionized and evaporated off the cloud in the top--triggered run. However, the volume from which most of the star--forming gas is drawn in all runs is very similar, while exactly which gas particles form stars clearly changes significantly between runs, as shown by the complex admixture of red and black traces over most of the central regions of the clouds. In Figures \ref{fig:trig_sink1} and \ref{fig:trig_sink2} we repeat this exercise at the level of individual sinks. The left panels of both figures show the traces in the top run of gas particles which form particular sinks. The right panels show the traces of the same gas particles in the control run, with particles that are involved in star formation in the control run plotted in black, and those not involved plotted in red. Figure \ref{fig:trig_sink1} shows an object for which \emph{none} of the corresponding gas in the control run is involved in star formation while Figure \ref{fig:trig_sink2} shows an object on which only a small fraction ($\sim30\%$) of the corresponding gas in the control run is involved in star formation.\\
\indent We therefore define an even more conservative criterion for determining whether a star is triggered or aborted, i.e. one which will report the fewest numbers of such events: if less than half the material forming a given star in one of the triggered simulations is \emph{involved in star formation} in the control run, the star is counted as triggered, otherwise the star is counted as spontaneous. Conversely, if less than half the material forming a given star in the control run is involved in star formation in one of the ionized runs, that star is counted as aborted in that run. We term this the `involvement' method. The same--seed and same--star techniques focus largely on whether or not particular identifiable objects form, whereas this method is more general and aims to identify whether particular collections of gas particles form stars in different runs without inquiring which, or how many, stars they contribute to. Both the objects shown in Figures \ref{fig:trig_sink1} and \ref{fig:trig_sink2} are defined as triggered by this criterion. In Figure \ref{fig:trig}, we show the results of applying this analysis to the top and bottom runs, where triggered stars are marked in red and spontaneously--formed stars in black.\\
\indent In Figures \ref{fig:trig_sink1} and \ref{fig:trig_sink2} we show examples, both drawn from the top and control runs, of sinks whose formation we regard as triggered.\\
\indent The distribution of stars in both ionized simulations is very similar to that in the control run, with most of the stars forming in a filament lying roughly along the $y$--axis. In both ionized runs, stars in the filament are a mixture of triggered and spontaneous objects using the involvement criterion. This reveals that what is happening in the central filament is not merely that the same gas is forming different permutations of stars in all three runs, but that \emph{different gas} is involved in star formation in the ionized runs. The top run contains 196 stars of which 103 are triggered and 93 are spontaneous, whereas the bottom run contains 206 stars of which 39 are triggered and 167 are spontaneous. The majority of triggered objects are mixed in with the spontaneously--formed ones in the central filament, making the two groups impossible to distinguish spatially. Only in the outlying areas of the triggered clusters -- the top--right of the top run and the bottom--right of the bottom run -- are there any outstanding groups of triggered stars, and their numbers are small. We also compared the velocities of the triggered and spontaneous populations parallel and perpendicular to a line--of--sight along the $z$--axis, since these are quantities an observer could measure. We find that there is nothing in these quantities to distinguish the triggered and spontaneous populations.\\
\indent We also applied the involvement criterion to identify aborted stars. In Figure \ref{fig:abort}, we plot the locations of stars in the control run which also form in the top (left panel) and bottom (right panel) simulations as black dots, whereas those that are aborted in each run are plotted as red dots. We see, again, that the objects whose fate has been changed by feedback are spatially mixed with those which form regardless. We conclude from the distributions of triggered, aborted and spontaneously--formed stars in the central filament that feedback agitates the gas in this region so that different parcels of material go to form stars in each run, although most of the star--forming gas is drawn from roughly the same volume, as shown by Figures \ref{fig:trace_control} and \ref{fig:trace_bound}. However, triggering and abortion in this region roughly cancel each other out, so that the total stellar mass and total numbers of stars formed is nearly the same in all threes simulations. The greater number of triggered and aborted objects in the top run is a result of the greater agitation of the gas in the central star--forming filament, due to the source in this run being closer to the filament and to the lack of intervening gas that could shield the filament from the radiation or the shocks driven by photoevaporation. Although it is possible to identify triggering and abortion of stars in the filament in the simulations, since these processes cancel each other out and the result is very nearly the same total stellar mass, number of stars and stellar mass function, the results of this analysis are of little importance from the point of view of the global properties of the cloud and its star cluster.\\
\indent It might be expected that stars whose formation has been triggered should be located close to the ionization front and to be moving along with it or not far behind it. In the top run, the ionization front rapidly reaches the central star--forming filament, after which point, almost all stars in the cloud are perforce located near to the front. However, the ionization front also moves into gas in the top right of the cloud, above the dense filament, triggering the formation of several stars. In Figure \ref{fig:trigtopright}, we show a time series of of column--density maps with the positions and velocities of the stars overlaid as white arrows, and the velocities of randomly--chosen gas particles overlaid as blue arrows. In the earliest frame, the velocities of the stars are clearly correlated with the motion of the gas and with the direction to the radiation source -- the stars were formed in, and  are moving along with, the dense gas behind the ionization front, which is travelling from left to right. This is still largely true in the middle panel, but the situation is now complicated, particularly in the stars that have formed in the dense filament whose formation and velocities are mostly unconnected with the ionization front. In addition, even amongst the stars outside the filament, most of which are triggered, dynamical interactions have begun to erase the stars' memories of their velocities at formation and the correlation with the gas velocity field is weaker. The correlation has been further eroded in the right panel of Figure \ref{fig:trigtopright}. In Figure \ref{fig:dcomp_top}, we plot the density PDF of the top and control runs, confined to the region shown in Figure \ref{fig:trigtopright}. The result is very similar to Figure \ref{fig:dens}, demonstrating that, although some triggering is taking place in this region, the evolution is still largely dominated by the dense filament formed by the turbulence.\\
\indent We observe a similar phenomenon in the bottom right of the bottom run. In Figure \ref{fig:trigbotright}, we show a second time series of column--density maps from the bottom run where the ionization front is moving though the low--density gas. In the first image, a tight, strongly--bound cluster of six stars is induced to form by the ionization front, all acquiring velocities close to that at which the gas behind the ionization front is moving. The second panel shows that two of the stars in the triggered cluster have been ejected by dynamical interactions and are now moving in a direction almost perpendicular to the motion of the ionization front, while a third star is ejected in a direction deeper into the front. These three objects are no longer bound to the small triggered cluster. We again plot in Figure \ref{fig:dcomp_bot} density PDFs from the bottom run (red) and the control run (black) confined to the region shown in Figure \ref{fig:trigbotright}. This time, we see that the density enhancement in this region, far from the densest gas in the simulation, is significantly higher than that seen for the whole cluster in Figure \ref{fig:dens}. Figures \ref{fig:trigtopright} and \ref{fig:dcomp_top} show that proximity to the ionization front does not necessarily imply that stars have been induced to form, since the front may run into a region where stars are forming anyway, and that the evolution of the gas such regions of the cloud is predetermined by the seed turbulence. Figures \ref{fig:trigtopright} and \ref{fig:trigbotright} show that dynamical interactions between triggered stars may erase their memory of moving along with the ionization front. The ionization front may induce the formation of small, tightly--bound clusters, in which all the stars are moving along with the front, but such clusters may be unstable, ejecting some of their members in directions uncorrelated with the motion of the ionization front. However, Figures \ref{fig:trigbotright} and \ref{fig:dcomp_bot} show that external triggering may dominate the evolution of some locations by sweeping up regions of low--density gas in which self--gravity has not yet asserted itself.\\
\section{Discussion}
\begin{figure}
\includegraphics[width=0.45\textwidth]{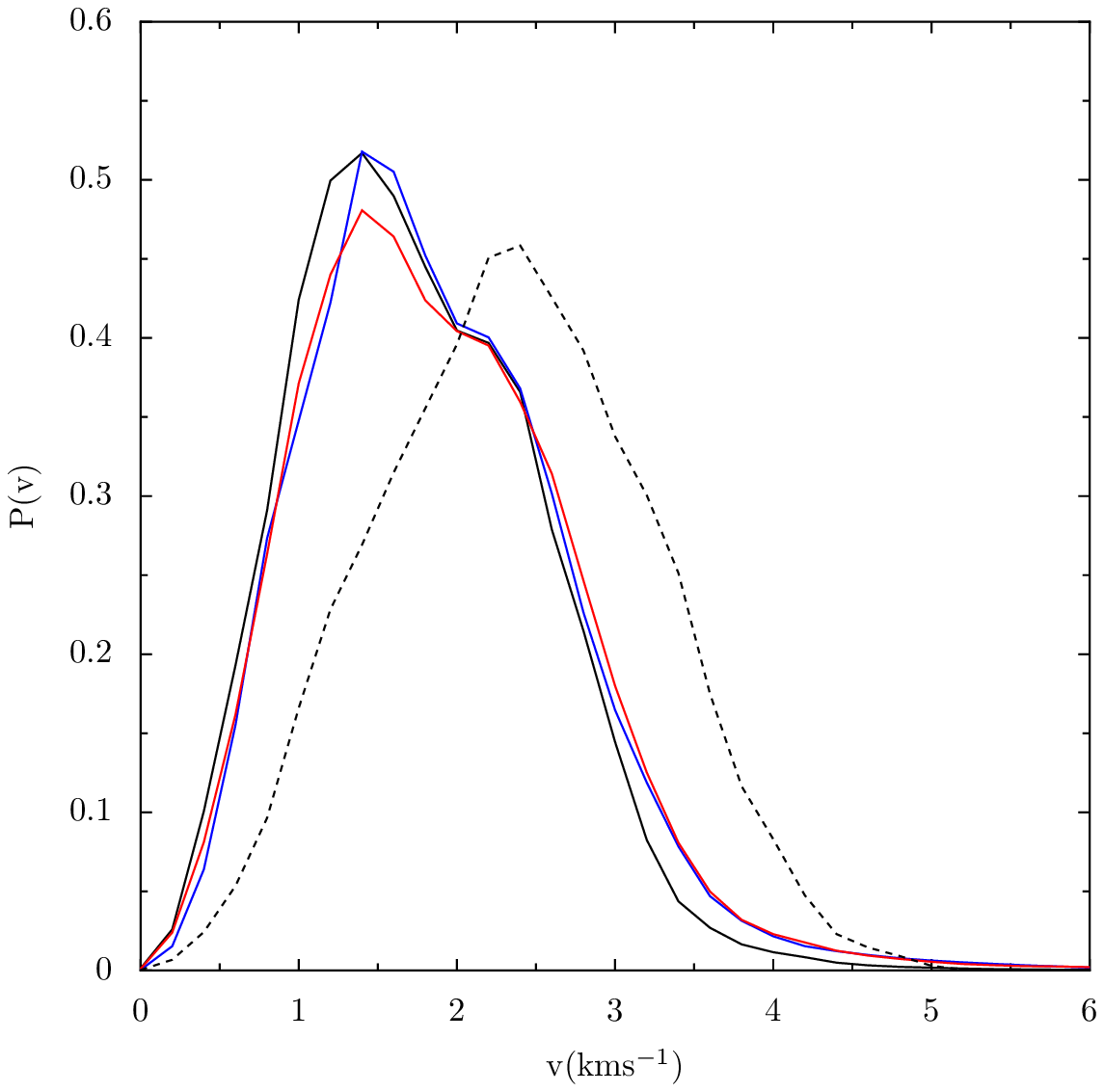}
\label{fig:vel}
\caption{Velocity probability distribution functions for the initial conditions of the simulations (black dashed line) and the final states of the control run (black solid line), the top run (red solid line) and the bottom run (blue solid line).}
\end{figure}

The purpose of this study was to see if external triggering of star formation in a molecular cloud has any statistical effect on the observable properties of the stars formed, such as their mass function, spatial distribution or velocities. We found that feedback had little effect on the numbers of stars formed and the mass functions in our control and triggered runs were statistically indistinguishable. The star formation rate in all three simulations is large controlled by the interplay between turbulence and gravity, which forms the dense central filament in which most of the stars form and which the ionization fronts driven by the external O--stars struggle to influence. The shocks driven by photoevaporation of the periphery of the cloud are able only to perturb slightly the density and velocity field of the gas within this central region. Although this changes exactly which parcels of gas become involved in star formation and which do not, so that some objects are effectively triggered and others are aborted, these effects cancel out in this region and the total mass of gas involved in star formation is virtually unchanged. Since the gas densities and velocities in the region of the cloud where most of the stars form are largely unaffected by feedback, there is also no change to the mass function of stars produced. This work suggests that is difficult for an external ionization source to influence the numbers or types of stars that a bound turbulent cloud is going to form. Work by \cite{2009MNRAS.397..232B} and \cite{2009MNRAS.398...33P} shows that \emph{internal} feedback from \emph{low--mass} stars has a much stronger effect on fragmentation and on the IMF.\\
\begin{figure*}
     \centering
     \subfigure[Column density map viewed along the $z$--axis of the bottom right corner of the bottom run at 0.531 Myr.]{\includegraphics[width=0.31\textwidth]{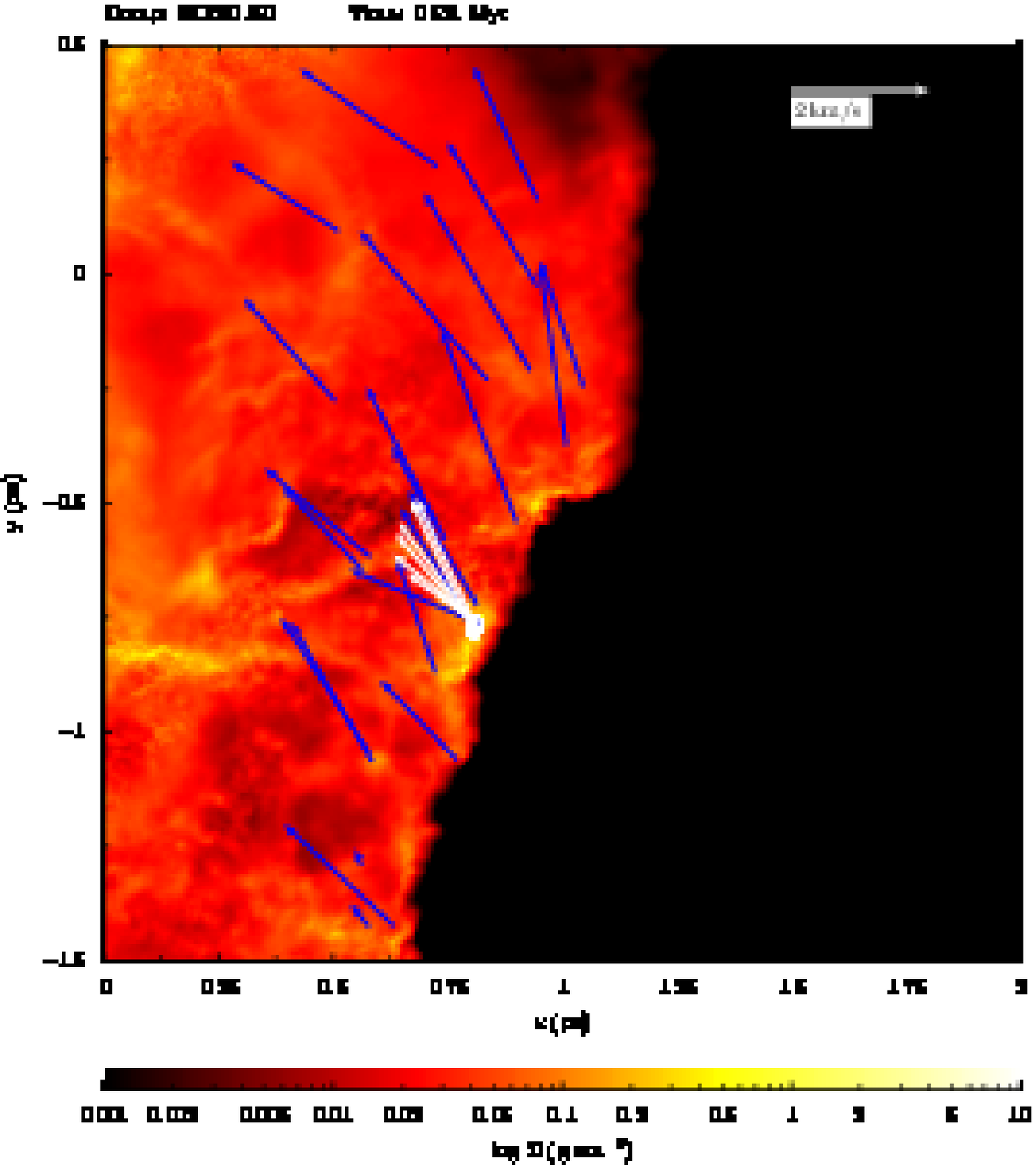}}     
     \hspace{.1in}
     \subfigure[Column density map viewed along the $z$--axis of the bottom right corner of the bottom run at 0.595 Myr.]{\includegraphics[width=0.31\textwidth]{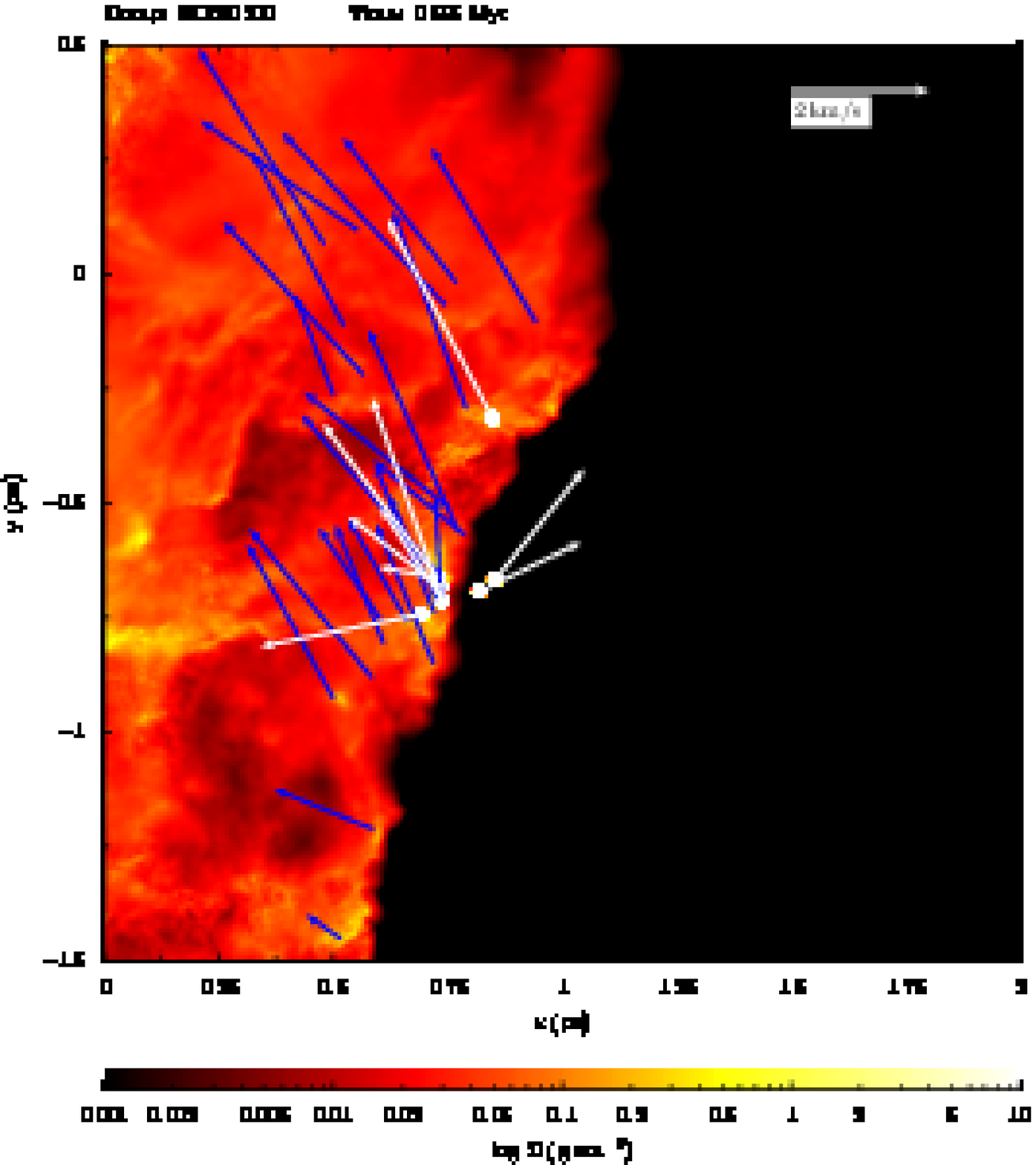}}
     \hspace{.1in}
     \subfigure[Column density map viewed along the $z$--axis of the bottom right corner of the bottom run at 0.651 Myr.]{\includegraphics[width=0.31\textwidth]{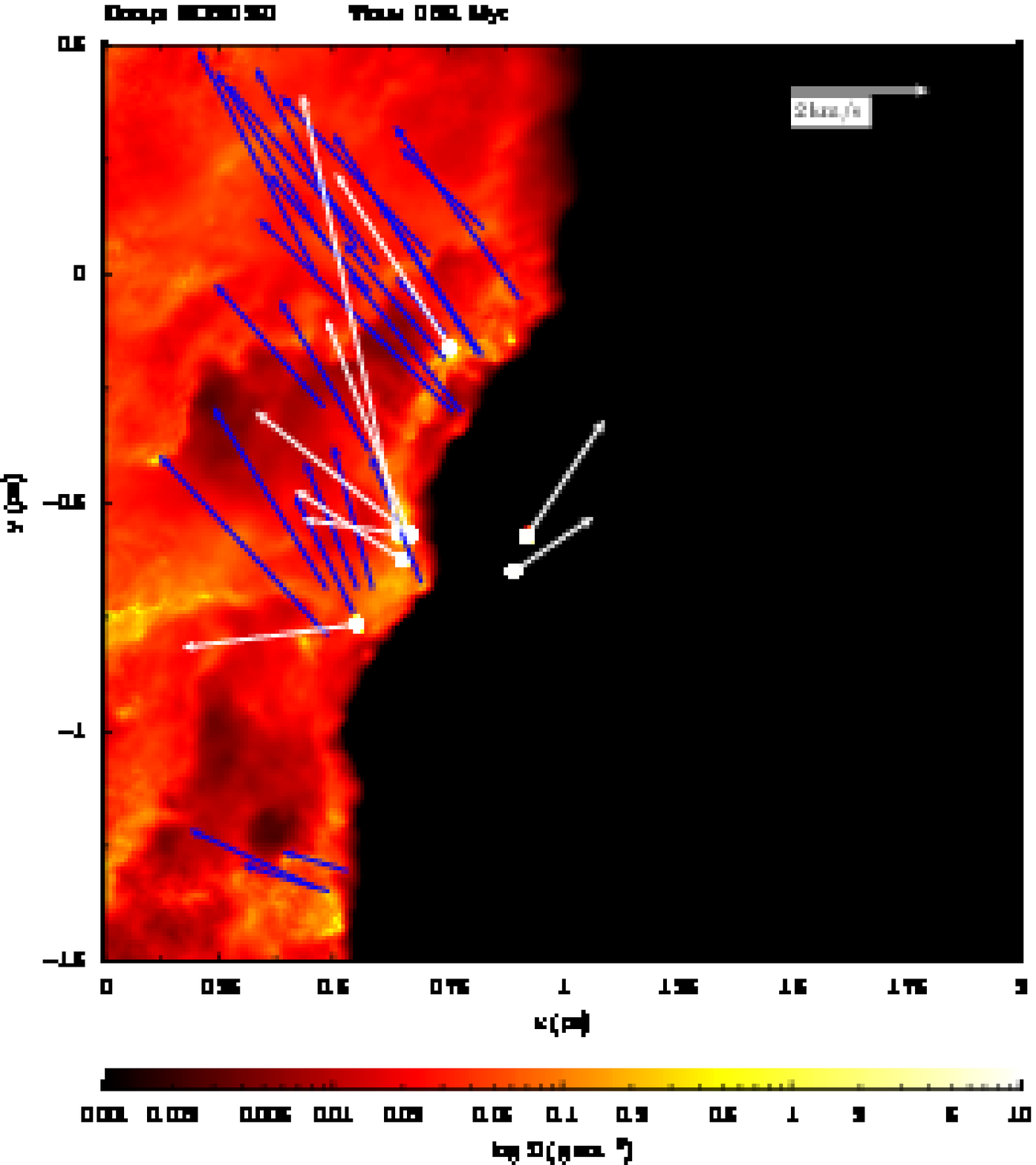}}
     \caption{Motions of the stars (white arrows) and randomly--selected gas particles (blue arrows) in the bottom right corner of the bottom run at three different epochs.}
   \label{fig:trigbotright}
\end{figure*}   
\indent Although we can identify triggered and spontaneously--formed stars by reference to the detailed output of our simulations, most of them cannot be identified as such by observing their positions or velocities. In particular the locations and velocities of those stars that are triggered are not necessarily related to the location and velocity of the ionization front. Overall, we find that the characteristics of the star formation are not strongly influenced by feedback. However, feedback does strongly modify the appearance of the cloud and, as shown in Figures \ref{fig:trigbotright} and \ref{fig:dcomp_bot}, and to a lesser extent, Figure \ref{fig:trigtopright}, may be able to dominate the evolution of lower density parts of the cloud. Even in such regions, the distribution of stars and gas is complicated by dynamical interaction between the stars which swiftly erases the correlation in position and velocity between triggered stars and the ionization front.\\
\indent The initial conditions used in our simulations are that of a bound turbulent molecular cloud. The control run, in which there is no feedback, is therefore governed by the interplay of the turbulent velocity field and the gas self--gravity. In Figure \ref{fig:vel} we plot the velocity probability distribution functions in the three simulations and compare them to that of the initial conditions. We see that the peak in the distributions has fallen from $\approx2.5$ km s$^{-1}$ initially to $\approx1.5$ km s$^{-1}$, due to the dissipation of turbulent kinetic energy, and that both feedback runs exhibit a broader high--velocity tail and slightly less very low--velocity material. Overall, however, the effect of external feedback on the velocity field is clearly slight and the driving of turbulence by feedback is weak.\\
\indent It is likely that an ionizing source with a significantly higher photon luminosity, or one which was placed closer to our model clouds may produce a stronger triggering (or disruption effect), since a higher photon flux at the cloud surface and a faster photoevaporation rate would result. However, as explained in Section 2, the photon luminosity we have used is appropriate for a small star cluster of mass comparable to the mass of our clouds and likely to have a radius comparable to the chosen initial separation between the source and the cloud edge. We cannot therefore increase the incident photon flux significantly without invoking a much brighter (and, by implication, more massive) radiation source, or moving the source much closer to the target clouds. In either case, it would be unrealistic to ignore the \emph{gravitational} influence such a source should have on the clouds. We therefore consider the photon flux resulting from our choice of source and separation to be towards the high end of what is realistic for the irradiation of the clouds considered in this paper.\\
\indent The use of different target clouds may also lead to stronger or weaker effects of feedback. The most important factor restraining the influence of photionization in these calculations is the high density gas in the central region of the cloud where most of the star formation takes place regardless of feedback. Clouds in which the gas density is lower are likely to feel the effects of photionization more strongly. For a cloud with mass $M$ and initial radius $R$, the initial gas density $\rho_{0}\sim M/R^{3}$. If we insist that the virial ratio is constant and assume that the gas thermal energy is negligible in comparison to the turbulent kinetic energy, $M/R\sim v_{\rm RMS}^{2}$, where $v_{\rm RMS}$ is the root--mean--square turbulent velocity. The maximum density $\rho_{\rm MAX}$, which is likely to describe the gas where the stars begin to form, will be generated by shocking in the turbulent flows. If the shocks are approximately isothermal and the typical sound speed in the quiescent gas is $c_{\rm s}$, $\rho_{\rm MAX}\sim\rho_{0}(v_{\rm RMS}/c_{\rm s})^{2}$. An increase in the initial cloud radius of a factor of two would then decrease the initial density by a factor of eight and the maximum density by a factor of sixteen. This may be sufficient to allow feedback to have a greater influence on the cloud. However, it is not clear whether this would lead to more destruction of the cloud, or more triggering. In addition, our simulations in \citep{2007MNRAS.377..535D} do have a very much lower gas density than those presented here and we found that the impact of feedback on star formation in this cloud was also rather modest. It is also possible increasing the relative velocity of the photoionization--driven shocks to the turbuent shocks, either by lowering $v_{\rm RMS}$ or by lowering the gas density so that mass--loading does not slow the feedback--generated shocks so much, would allow feedback to have a greater influence on the cloud but, again, it is not clear whether this influence would be more positive of more negative. A full evaluation of these questions demands a parameter study, which we defer to later work.
\begin{figure}
\includegraphics[width=0.45\textwidth]{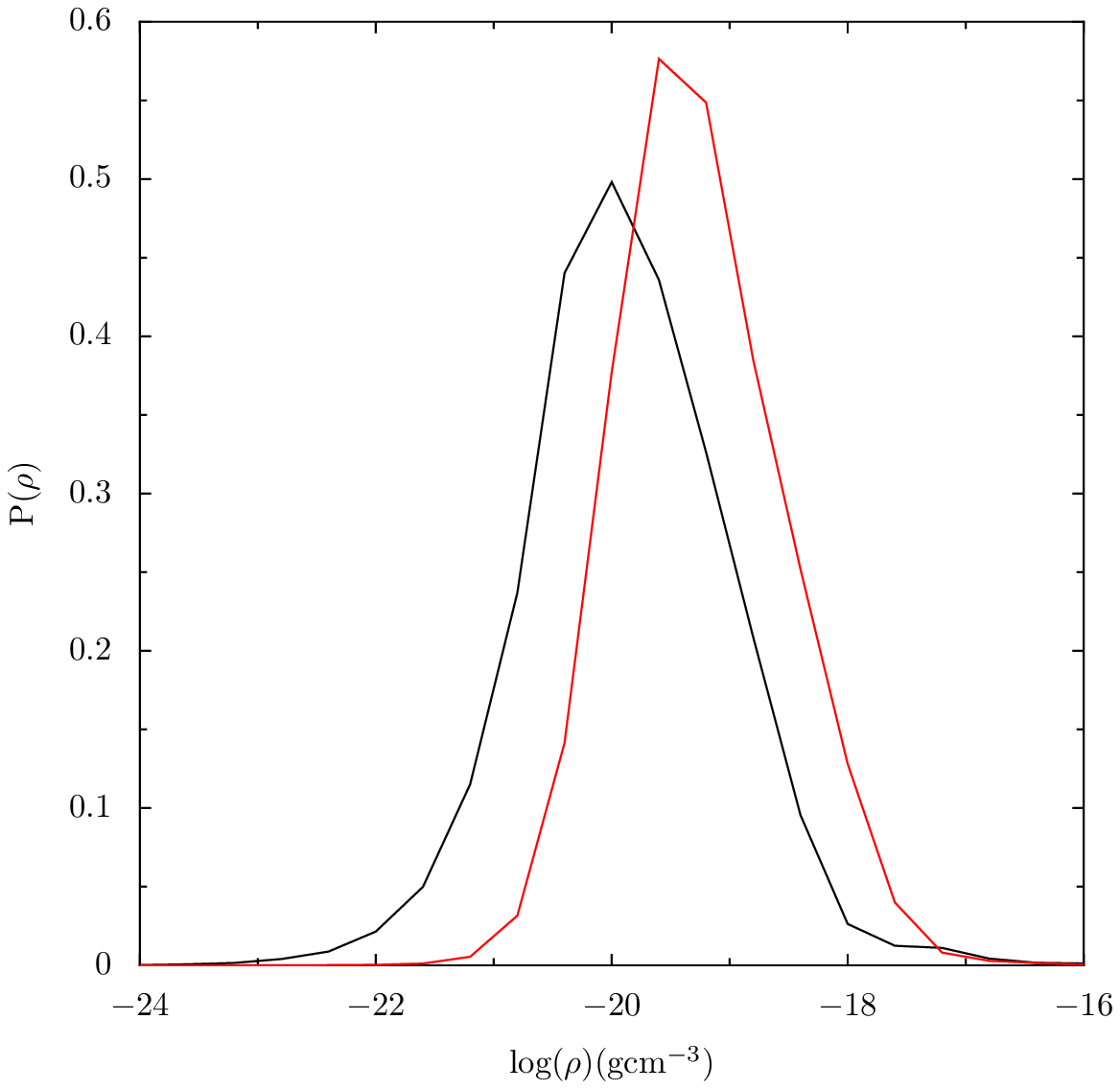}
\caption{Comparison of the density PDFs of the bottom run (red) and the control run (black) in the region shown in Figure \ref{fig:trigbotright} at a time of 0.595Myr.}
\label{fig:dcomp_bot}
\end{figure}
\section{Conclusions}
We find that the effect of external photoionizing irradiation by an O--star on our bound molecular cloud is modest. Although the evaporation of gas alters the morphology of the gas quite strongly in some regions of the cloud, the effect on the rate and efficiency of star formation is small. The effects of feedback on the stellar mass functions were also statistically negligible. We also find that the morphology of the stellar clusters produced in our feedback runs are very similar to that in the control run, and that the majority of stars whose formation was triggered cannot be distinguished from their spontaneously--formed siblings by their positions or velocities with respect to the ionization front. Only in the lower--density peripheral regions of the cloud, near the ionizing sources, do distinct populations of triggered stars form.\\
\indent Feedback has modest effects on the density and velocity probability density functions in our clouds. Instead, the turbulent velocity field and the density field it generates with the assistance of gravity are the dominant agencies controlling the tempo and mode of star formation. This need not aways be the case and further studies are required to evaluate the dependence of this conclusion on, for example, the turbulent velocity field, the boundedness of the cloud and luminosity of the photoionizing radiation source.

\section{Acknowledgements}
We thank the anonymous referee for insightful suggestions which significantly improved the paper.

\bibliography{myrefs}

\label{lastpage}

\end{document}